\documentclass[onecollarge,natbib,referee]{svjour2}
\bibpunct{[}{]}{;}{n}{}{,} 
\smartqed  
\usepackage{amsmath,revsymb,graphicx,amssymb,youngtab}
\newcommand{\kt}[1]{\ensuremath{|{#1}\rangle}}
\newcommand{\br}[1]{\ensuremath{\langle {#1}|}}
\newcommand{\bk}[2]{\ensuremath{\langle {#1}|{#2}\rangle}}
\newcommand{\ykt}[1]{\ensuremath{\left| \Yvcentermath1 {#1} \Yvcentermath0 \right\rangle}}

\newcommand{\rykt}[1]{\ensuremath{\left\| \Yvcentermath1 {#1} \Yvcentermath0 \right\rangle}}
\newcommand{\rybr}[1]{\ensuremath{\left\langle\Yvcentermath1 {#1} \Yvcentermath0 \right\|}}
\newcommand{\HS}{\mathcal{H}}
\newcommand{\SHS}{\mathcal{S}}
\newcommand{\KHS}{\mathcal{K}}
\newcommand{\pp}[1]{\ensuremath{{\lfloor #1 \rfloor }}}

\newcommand{\bep}{\ensuremath{\boldsymbol{\epsilon}}}

\newcommand{\mrep}[1]{\mathcal{M}^{#1}}
\journalname{Few-Body Systems}
\begin{document}

\title{One-Dimensional Traps, Two-Body Interactions, Few-Body Symmetries}

\subtitle{II.~$N$ Particles}

\titlerunning{One, Two, Few: II.~$N$}  

\author{N.L.\ Harshman}


\institute{N.L.\ Harshman \at
              Department of Physics\\
              American University\\
              4400 Massachusetts Ave. NW\\
              Washington, DC 20016-8058 USA\\
              Tel.: +1-202-885-3479\\
              Fax: +1-202-885-2723\\
              \email{harshman@american.edu}           
}

\date{Received: date / Accepted: date}

\maketitle

\begin{abstract}
This is the second in a pair of articles that classify the configuration space and kinematic symmetry groups for $N$ identical particles in one-dimensional traps experiencing Galilean-invariant two-body interactions. These symmetries explain degeneracies in the few-body spectrum and demonstrate how tuning the trap shape and the particle interactions can manipulate these degeneracies. The additional symmetries that emerge in the non-interacting limit and in the unitary limit of an infinitely strong contact interaction are sufficient to algebraically solve for the spectrum and degeneracy in terms of the one-particle observables. Symmetry also determines the degree to which the algebraic expressions for energy level shifts by weak interactions or nearly-unitary interactions are universal, i.e.\ independent of trap shape and details of the interaction. Identical fermions and bosons with and without spin are considered. This article analyzes the symmetries of $N$ particles in asymmetric, symmetric, and harmonic traps; the prequel article treats the one, two and three particle cases. 
\end{abstract}

\section{Introduction to Part II}
\label{intro}

This is the second in a pair of articles that classifies the symmetries of a model system of identical particles trapped in one-dimension and interacting via Galilean-invariant two-body interactions. The first article motivated the study of this system and its symmetries and it considered examples and applications with one, two and three particles. This article formalizes and extends these results to the case of $N$ particles, with multiple examples for $N=4$.

The Hamiltonian under study is 
\begin{equation}
\hat{H}^N = \hat{H}^N_0 + \hat{V}^N,
\end{equation}
where 
\begin{equation}
\hat{H}^N_0 = \sum_{i=1}^N \hat{H}^1_i 
\end{equation}
is the non-interacting Hamiltonian. It is the sum of identical one-particle Hamiltonians $\hat{H}^1_i$ that each include the one-particle trapping potential $\hat{V}^1(\hat{Q}_i)$. The interacting Hamiltonian $\hat{H}^N$ includes the sum of identical pairwise interaction potentials
\begin{equation}
\hat{V}^N = \sum_{i<j}^N \hat{V}_{ij}.
\end{equation}
The specific case of contact interactions is treated in detail, but many results hold for \emph{any} Galilean invariant interaction potential $\hat{V}_{ij}=V^2(|\hat{Q}_i - \hat{Q}_j|)$.

The goal is to classify the symmetries of $\hat{H}^N_0$ and  $\hat{H}^N$ for asymmetric, symmetric and harmonic traps. Two kinds of symmetries are considered: configuration space symmetries, which are realized as orthogonal transformations of configuration space ${\bf q} \in \mathcal{Q}^N=\mathbb{R}^N$, and kinematic symmetries, which are the full group of all unitary transformations that commute with $\hat{H}^N_0$ or $\hat{H}^N$. Despite their importance and intrinsic interest, neither this article nor its prequel consider dynamical (a.k.a.\ spectrum-generating) symmetries.

Key to the methods and results of this article is the assumption that each one particle Hamiltonian $\hat{H}^1_i$ has a discrete, non-degenerate energy spectrum $\sigma_1=\{\epsilon_0, \epsilon_1,\epsilon_2,\ldots\}$. The configuration space symmetry $\mathrm{C}_1$ is either trivial $\mathrm{C}_1\sim\mathrm{Z}_1$ for asymmetric traps or it is parity $\mathrm{C}_1\sim\mathrm{O}(1)$ for symmetric traps. The kinematic symmetry group $\mathrm{K}_1$ is time translation $\mathrm{K}_1 \sim \mathrm{T}_t$ for asymmetric traps, parity and time translation $\mathrm{K}_1\sim \mathrm{O}(1) \times \mathrm{T}_t$ for symmetric traps, or the unitary group and time translation for the harmonic trap $\mathrm{K}_1 \sim \mathrm{U}(1) \times \mathrm{T}_t$. See Sections 2 of the previous article for more details about one particle symmetries.

The results for $N=2$ and $N=3$ in the previous article could be established by direct calculation and enumeration. However, that is not practical for higher $N$. The goal is to develop algebraic methods that can be implemented on a computer, but this requires a degree of formality and abstraction that some physicists may find unfamiliar or unappealing, i.e.\ the `Gruppenpest'. To try and bridge that gap, the second section of this article starts out with an overview of the permutation group and its representations. A particular class of representations called permutation modules turn out to be intuitive and useful, especially when the tools of state permutation symmetry and the double tableau basis are employed. This section also discussed how to apply these results to the case where the $N$ identical particles are fermions or bosons and have an internal structure that does not participate in the interaction, like spin.

The third section considers the non-interacting Hamiltonian $\hat{H}^N_0$ and shows that the configuration space symmetry group $\mathrm{C}_N^0$ and the kinematic symmetry group $\mathrm{K}_N^0$ are always larger groups than the following minimal constructions
\begin{equation}
\mathrm{C}_N^0 \subseteq \mathrm{P}_N \ltimes \mathrm{C}_1^{\times N}\ \mbox{and}\  \mathrm{K}_N^0 \subseteq \mathrm{P}_N \ltimes \mathrm{K}_1^{\times N}.
\end{equation}
In words, these symmetry groups are at least the semidirect product of the particle permutation group $\mathrm{P}_N$ acting on the direct product of $N$ copies of the single particle symmetry group $\mathrm{C}_1$ or $\mathrm{K}_1$. The previous article developed this structure by analysis of two and three particle systems, where the group elements can be easily explicitly enumerated. In this article, the general result is established by using representation theory for $\mathrm{P}_N \sim \mathrm{S}_N$ applied to permutation modules. The degeneracies of the non-interacting spectrum $\sigma_N^0$ must be explained by the dimensionality of the irreducible representations of $\mathrm{K}_N^0$. When $\mathrm{C}_N^0$ or $\mathrm{K}_N^0$ is larger than the minimal construction, then there is either an emergent few-particle symmetry or an accidental symmetry. Table 1 summarizes results for $2$, $3$ and $4$ non-interacting particles.

The fourth section shows that there is also a minimal construction for the configuration space symmetry group $\mathrm{C}_N$ and the kinematic symmetry group $\mathrm{K}_N$ of the interacting Hamiltonian $\hat{H}^N$:
\begin{equation}
\mathrm{C}_N \subseteq \mathrm{P}_N \times \mathrm{C}_1\ \mbox{and}\  \mathrm{K}_N \subseteq \mathrm{P}_N \times \mathrm{K}_1.
\end{equation}
The interactions break individual particle symmetries and the non-interacting energy levels are split into irreducible representations of the smaller group $\mathrm{K}_N$. Table 2 summarizes these results for $2$, $3$ and $4$ interacting particles. This reduction of symmetry can be exploited to make approximation schemes like exact diagonalization scale more efficiently for the same level of accuracy and to find algebraic results for level-splitting under weak perturbations.

The fifth section considers the unitary limit of the contact interaction, with Hamiltonian denoted $\hat{H}^N_\infty$. In this limit, the ordering of particles becomes a good quantum number and a new kinematic symmetry emerges for finite-energy states
\begin{equation}
\mathrm{K}^\infty_N \sim \mathrm{P}_N \times \mathfrak{O}_N \times \mathrm{K}_1.
\end{equation}
The additional symmetry $\mathfrak{O}_N \sim \mathrm{S}_N$ is called ordering permutation symmetry, and it provides and alternate set of observables that can be used to analyze how the $N!$-fold degenerate energy levels split into less degenerate levels and bands in the `near unitarity' limit.

Throughout the article the question is asked, what results are universal? Specifically, when is there enough symmetry for the spectrum of $\hat{H}^N$ to be calculated from the properties of the one-particle system? Which properties are required for which approximations, and what can be said without specific knowledge of the trap shape or the interaction? How does this change with increasing particles or spin components? The short answer is that the non-interacting limit and the unitary limit of the contact interaction are algebraically universal for any $N$. Only the single particle spectrum $\sigma_1$ is required in order for the machinery of group representation theory to construct a complete set of commuting observables for those two limits. For other interactions, the interplay of the trap shape, number of particles, and specific interaction determine whether algebraically universal expressions exist for properties like level splitting of multicomponent particles under weak perturbations or near-unitary perturbations.

\section{The Symmetric Group}

The previous article discusses one, two and three particles and group representation theory techniques is employed to make well-known or intuitive results seems reasonable and inevitable. For two and three particles, it is relatively easy to achieve results by manual calculation or enumeration; the biggest composition subspaces for three particle are only six-dimensional. When we get to four particles, the possibility for 24-fold degenerate energy levels, even without accidental or emergent symmetries, encourages us to develop more sophisticated techniques. At the core of these techniques is the theory of the symmetric group and its representations. This section develops the necessary framework to extend results for two and three particles to four and more.

The Hamiltonian is invariant under particle permutations $\mathrm{P}_N$ for every kind of trap and for any Galilean-invariant interactions. Therefore the configuration space symmetry group and the kinematic symmetry group have  the abstract group $\mathrm{S}_N$ as a subgroup\footnote{The distinction between $\mathrm{P}_N$ as a physical symmetry and $\mathrm{S}_N$ as the abstract group is useful because state permutation symmetry $\mathfrak{P}_\pp{\nu}$ and ordering permutation symmetry $\mathfrak{O}_N$ are also isomorphic to a symmetric group.}. The properties of $\mathrm{S}_N$ and its irreducible representations (irreps) are well-known (c.f.\ \cite{ Hamermesh, Sagan, Chen, Ma}) and frequently applied in few-body physics. The first subsection establishes the local notation and definitions for $\mathrm{S}_N$ elements and irreps. Experts in the symmetric group could probably skip this subsection. Novices in group theory may find it useful to read subsection 1.4 from the previous paper, which introduces some notation and definitions for groups and their representations. The second subsection defines compositions and permutation modules, and the third discusses how to incorporate spin and spatial degrees of freedom by taking direct products of the spatial Hilbert space $\KHS$ and the spin Hilbert space $\SHS$.

\subsection{$\mathrm{S}_N$ Definitions}

Elements of $\mathrm{S}_N$ can be denoted by permutations $p=\{i_1 \ldots i_N\}$ or cycles $c=(ij\cdots k)$. For example, the permutation $p=\{1324\}$ and cycle $c=(23)$ realize the same element of $\mathrm{S}_4$. All elements with the same cycle structure form a conjugacy class. A partition of $N$ is a set of non-negative integers $[\mu]=[\mu_1\mu_2\ldots\mu_r]$ that sum to $N$. Denote the set of partitions of $N$ by $P(N)$. There is a conjugacy class of $\mathrm{S}_N$ for each partition $[\mu]\in P(N)$. As an example, the five partitions and conjugacy classes for $\mathrm{S}_4$ are:
\begin{itemize}
\item $[1111] \equiv [1^4]$: 4 one-cycles, i.e.\ the identity $e=()=\{1234\}$;
\item $[211] \equiv [21^2]$: 1 two-cycle and 2 one-cycles, also called transpositions, i.e.\ the six transpositions $(12) \equiv \{2134\}$, $(13)$, $(14)$, $(23)$ , $(24)$, and $(34)$;
\item $[22] \equiv [2^2]$: 2 two-cycles, i.e.\ the three disjoint, double transpositions $(12)(34)=\{2143\}$, $(13)(24)$, and $(14)(23)$;
\item $[31]$: 1 three-cycle and 1 one-cycle, i.e.\ the eight permutations $(123)=\{2314\}$, $(132)$, $(124)$, $(142)$, $(134)$, $(143)$, $(234)$, and $(243)$;
\item $[4]$: 1 four cycle, i.e.\ the six permutations $(1234)=\{2341\}$, $(1243)$, $(1324)$, $(1342)$, $(1423)$, $(1432)$.
\end{itemize}
For the symmetric group, an element and its inverse are in the same conjugacy class. All elements in a conjugacy class are even or odd depending on whether they can be generated by an even or odd number of transpositions.

For each partition of $N$ there is also an irreducible representation (irrep) of $\mathrm{S}_N$. These irreps form building blocks for other representations of $\mathrm{S}_N$, like the configuration space representation $\underline{O}(p)$ or the unitary Hilbert space representation $\hat{U}(p)$. Irrep labels can be depicted as Ferrers diagrams, i.e.\ $r$ rows of boxes with the $i$th row having $\mu_i$ boxes (also called Young diagrams).  The Ferrers diagrams for $N=4$ in order from least to greatest are 
\[
\yng(4),\ \yng(3,1),\ \yng(2,2),\ \yng(2,1,1),\ \yng(1,1,1,1).
\]
There is a canonical `lexicographic' ordering of irreps: the lowest partition $[N]$ is the one-dimensional, totally symmetric irrep and the highest partition $[1^N]$ is the one-dimensional, totally antisymmetric irrep. Other partitions correspond to multi-dimensional representations with mixed symmetry under permutations. The notation $[\mu]^\top$ indicates the conjugate irrep of $[\mu]$. A conjugate irrep is the partition of $N$ that has the Ferrers diagram with rows and columns reversed, e.g.\ $[31]^\top = [21^2]$,  and $[N]^\top = [1^N]$. Some partitions are self-conjugate, like $[2^2]$ for $N=4$ and $[31^2]$ for $N=5$.

The finite-dimensional vector space that carries the irrep $[\mu]$ is denoted $\mathcal{M}^{[\mu]}$. Denote the dimension of $\mathrm{S}_N$ irreps by $d[\mu]$. For $N=4$, these dimensions are $d[4]=1$, $d[31]=3$, $d[2^2]=2$, $d[21^2]=3$, and $d[1^4]=1$.  The irrep dimension $d[\mu]$ can be calculated using the Frobenius formula, the hook-length method, or by counting the number of standard Young tableaux\footnote{Standard Young tableaux are filled with numbers or labels that must increase to the right and to the bottom.} that are possible for a given Ferrers diagram, e.g.\ for irrep $[21^2]$ the standard Young tableaux are
\[
\young(12,3,4),\ \young(13,2,4),\ \young(14,2,3).
\]
The $\mathrm{S}_N$ irrep matrices $\underline{D}^{[\mu]}$ depend on the basis chosen for $\mrep{[\mu]}$ and there are several methods for selecting bases and generating these matrices. This article uses a standard Young tableau $Y$ to label a basis  $\kt{Y} \in \mrep{[\mu]}$ so that $p\in\mathrm{S}_N$ is represented  as 
\begin{equation}\label{eq:irrep}
\hat{U}(p)\kt{Y} = \sum_{Y'} \underline{D}^{[\mu]}_{YY'}(p) \kt{Y'}.
\end{equation}
The basis can always be chose so that the matrices $\underline{D}$ are real and orthogonal; a standard choice is the Yamanouchi basis convention~\cite{Hamermesh,Chen} in which the subgroup chain $\mathrm{S}_N \supset \mathrm{S}_{N-1} \supset \cdots \supset \mathrm{S}_2$ is diagonalized. For the totally symmetric representation $[N]$, the `matrices' $\underline{D}^{[N]}(p)=1$ are trivial for all $p\in\mathrm{S}_N$. For the totally antisymmetric representation, the matrices have the property $D^{[N]}(p)=\pi_p$, where $\pi_p$ is the signature of the permutation $p$, i.e.\ whether the permutation is even $\pi_p=1$ or odd $\pi_p=-1$.

\subsection{Compositions and Permutation Modules}

Consider a sequence of $N$ non-negative integers ${\bf n}=\langle n_1,n_2,\cdots,n_N \rangle$. The composition of ${\bf n}$ describes the numbers $n_i$ that appear in ${\bf n}$ and their degeneracies $\nu_i$ without regard to the particular sequence. One notation for a composition is $\pp{\nu} = \pp{0^{\nu_0} 1^{\nu_1} 2^{\nu_2} \ldots n^{\nu_n} \ldots}$, dropping terms with $\nu_i=0$ and omitting the exponent when $\nu_i=1$. For example, the sequence $\langle 2,0,1,4,1 \rangle$ has the composition $\pp{01^224}$. Instead of numbers, more general symbols can be used, e.g.\ the composition of a sequence of $5$ symbols $\langle \alpha \beta \beta \alpha \gamma\rangle$ is denoted $\pp{\nu}=\pp{\alpha^2\beta^2\gamma}$.  Note that $\sum_i \nu_i=N$ and the shape $[\nu]$ of a composition $\pp{\nu}$ must be a partition $[\nu]\in P(N)$, e.g.\ if $\pp{\nu}=\pp{01^224}$ then $[\nu]=[21^3]$ and if $\pp{\nu}=\pp{\alpha^2\beta^2\gamma}$ then $[\nu]=[2^21]$.

The set of all sequences ${\bf n}$ with the same composition $\pp{\nu}$ forms a basis for a representation space of $\mathrm{S}_N$ called a permutation module $M^{[\nu]}$. The action of $p\in\mathrm{S}_N$ on a basis sequence is
\begin{equation}
p \cdot \langle n_1,n_2,\cdots,n_N \rangle = \langle n_{p_1},n_{p_2},\cdots,n_{p_N} \rangle.
\end{equation}
The dimension $d\pp{\nu}$ of $M^{[\nu]}$, or equivalently the number of sequences with composition $\pp{\nu}$, depends only on the shape $[\nu]$ of the composition $\pp{\nu}$. The formula for $d\pp{\nu}$ is
\begin{equation}\label{eq:dim}
d\pp{\nu} = \frac{N!}{\nu_1! \nu_2! \cdots \nu_r!}.
\end{equation}
For example, if $\pp{\nu}=\pp{01^224}$ then $d\pp{\nu}=60$ and if $\pp{\nu}=\pp{\alpha^2\beta^2\gamma}$ then $d\pp{\nu}=30$.

The permutation module $M^{[N]}$ is built on sequences composed of a single symbol, like $\pp{\nu}=\pp{\alpha^N}$. It is equivalent to the lowest, symmetric irrep $\mrep{[N]}$. All other permutation modules are reducible with respect to $\mathrm{S}_N$:
\begin{equation}\label{eq:kostka}
M^{[\nu]} = \bigoplus_{[\mu]\leq [\nu]} K_{[\mu] [\nu]} \mrep{[\mu]}
\end{equation}
where $[\mu]\leq [\nu]$ means that the partition $[\mu]$ is lower than or equal to the composition shape $[\nu]$ in lexicographic ordering and $K_{[\mu] [\nu]}$ is the Kostka number describing the number of times the irrep $[\mu]$ appears in $M^{[\nu]}$. Methods for calculating the Kostka numbers are well-established, for example, using characters~\cite{Hamermesh}, using combinatoric methods~\cite{Sagan}, or using the intrinsic group of the composition~\cite{Chen}. As an example, for $N=4$ there are five types of permutation modules $M^{[\nu]}$, with the following reductions
\begin{eqnarray}\label{fourcat}
M^{[4]} &=& \mrep{[4]}\nonumber\\
M^{[31]} &=& \mrep{[4]} \oplus \mrep{[31]}\nonumber\\
M^{[2^2]} &=& \mrep{[4]}\oplus \mrep{[31]}\oplus \mrep{[2^2]}\nonumber\\
M^{[21^2]} &=& \mrep{[4]}\oplus 2\mrep{[31]}\oplus \mrep{[2^2]}\oplus \mrep{[21^2]}\nonumber\\
M^{[1^4]} &=& \mrep{[4]}\oplus 3\mrep{[31]}\oplus 2\mrep{[2^2]}\oplus 3 \mrep{[21^2]} \oplus  \mrep{[1^4]}
\end{eqnarray}
The first permutation module $M^{[4]}$ is the trivial, totally symmetric representation of $\mathrm{S}_4$. The second module $M^{[31]}$ is the reducible representation of $\mathrm{S}_4$ when the composition $\pp{\nu}$ has only one different symbol; it is called the defining representation of $\mathrm{S}_4$. Notice that the permutation modules $M^{[21^2]}$ and $M^{[1^4]}$ are not simply reducible; multiple copies of the same $\mathrm{S}_4$ irrep appear.

Different copies of the same irrep $\mrep{[\mu]}$ can be distinguished by semi-standard Weyl tableaux\footnote{Semi-standard Weyl tableaux are filled with numbers or labels that must increase to the bottom, but may be the same or increase to the right.}, e.g.\ for the composition $\pp{\nu}=\pp{\alpha\beta^2\gamma}$ with shape $[\nu]=[21^2]$, there are two Weyl tableaux $W$ with shape $[W]=[31]$
\begin{equation}
\Yvcentermath1 \young(\alpha\beta\beta,\gamma)\Yvcentermath0 \ \mbox{and}\ \Yvcentermath1\young(\alpha\beta\gamma,\beta)\Yvcentermath0
\end{equation}
corresponding to the two copies of $\mrep{[31]}$ in $M^{[21^2]}$. In this example, the two copies of $\mrep{[31]}$ that appear in the permutation module $M^{[21^2]}$ with composition $\pp{\nu}=\pp{\alpha\beta^2\gamma}$ are distinguished by how they transform under the exchange of the symbols $\alpha$ and $\gamma$. In addition to carrying a representation of $\mathrm{S}_4$, the permutation module $M^{[21^2]}$ carries a representation of symbol permutation symmetry (exchanging of $\alpha$ and $\gamma$) that is isomorphic to $\mathrm{S}_2$.

To generalize, permutation modules $M^{[\nu]}$ carry a representation of $\mathrm{S}_N$ realized by sequence permutations on a composition of symbols $\pp{\nu} = \pp{\alpha^{\nu_\alpha} \beta^{\nu_\beta} \gamma^{\nu_\gamma} \cdots}$. Except when $[\nu]=[N]$, this representation is reducible, but not necessarily simply reducible. Additionally, permutations module carries a representation of the symbol permutation symmetry, denoted  $\mathfrak{S}_\pp{\nu}$. Non-trivial symbol permutation symmetry occurs when there are symbols in the composition that occur the same number of times. For example, in the composition $\pp{\nu}=\pp{012^24}$, the symbols 0, 1 and 4 all appear once, so the symbol permutation symmetry is $\mathfrak{S}_\pp{\nu}=\mathrm{S}_3$. For the composition $\pp{\nu}=\pp{\alpha^2\beta^2\gamma}$, the symbol permutation symmetry is $\mathfrak{S}_\pp{\nu}=\mathrm{S}_2$ because $\alpha$ and $\beta$ can be exchanged. Symbol permutation symmetry can be used to distinguish between different copies of the same $\mathrm{S}_N$ irreps $\mrep{[\mu]}$ that appear in $M^{[\nu]}$. Denote the irreps of $\mathfrak{S}_\pp{\nu}$ by $\{\nu\}$. The interplay between sequence permutation symmetry and symbol permutation symmetry has multiple applications in the subsequent sections.

For any $N$, when the composition shape is $[\nu]=[1^N]$, the module $M^{[1^N]}$ carries the regular representation of $\mathrm{S}_N$: each irrep space $\mrep{[\mu]}$ appears as many times as the dimension $d[\mu]$ of the irrep and the total dimension of $M^{[1^N]}$ is $d\pp{\nu}=N!$. The regular representation also carries a representation of symbol permutation symmetry $\mathfrak{S}_{\pp{\nu}}$ that is isomorphic to $\mathrm{S}_N$, i.e. $\{\nu\}=[1^N]$ also. This double symmetry plays a special role in the case of the unitary limit of the contact interaction, as shown in subsection 3.2.1.

One basis for permutation modules is provided by the basis of sequences $\langle {\bf n} \rangle$. Alternatively, a basis for the permutation module $M^{[\nu]}$ can be labeled by the set of vectors $\kt{W\,Y}$, where $W\in\pp{\nu}$ are all the Weyl tableaux possible for the composition $\pp{\nu}$ and  $Y \in [W]$ are all the Young tableaux for possible for the Ferrers diagram with shape $[W]$. Elements of the sequence permutation group $p\in\mathrm{S}_N$ are realized by operators $\hat{U}(p)$ that mix basis vectors with the same $W$ and different $Y \in [W]$. 
\begin{equation}
\hat{U}(p) \kt{W \, Y} = \sum_{Y\in[W]} \underline{D}^{[W]}_{YY'}(p)\kt{W\,Y'}.
\end{equation}
For modules that are not simply reducible, elements of the symbol permutation group $\mathfrak{p}\in\mathfrak{S}_{\pp{\nu}}$ mix basis vectors with the same $Y$ and different $W \in \{\nu\}$:
\begin{equation}
\hat{U}(\mathfrak{p}) \kt{W \, Y} = \sum_{W\in\{\nu\}} \underline{D}^{\{\nu\}}_{WW'}(\mathfrak{p})\kt{W'\,Y}.
\end{equation}
The double tableaux basis are eigenvectors of conjugacy class operators constructed from the group algebra of $\mathrm{S}_N$ and $\mathfrak{S}_{\pp{\nu}}$ and from their canonical subgroup chains $\mathrm{S}_N \supset\mathrm{S}_{N-1} \supset \cdots \supset \mathrm{S}_2$ and $\mathfrak{S}_N \supset\mathfrak{S}_{N-1} \supset \cdots \supset \mathfrak{S}_2$. The explicit construction of these operators and determination of their eigenvalues is given in \cite{Chen, Ma}, and examples for two and three particles are in the previous article. Class operators built from transpositions are applied in \cite{Guan2009, Ma2009, Fang2011} to analyze the model Hamiltonian for multicomponent fermions and fermion-boson mixtures. They can be efficiently implemented using standard computational algebra programs, but their details are not required for the results of this article.

\subsection{Symmetrization of Identical Particles with and without Spin}

The $N$-particle Hamiltonian (interacting or non-interacting) always has $\mathrm{P}_N$ symmetry, and so the spatial Hilbert space $\KHS$ can be decomposed into subspaces corresponding to irreps $[\mu]\in P(N)$
\begin{equation}
\KHS = \bigoplus_{[\mu]\in P(N)} \KHS^{[\mu]}.
\end{equation}
For trapped particles, each subspace $\KHS^{[\mu]}$ is isomorphic to an infinite tower of irreps spaces $\mrep{[\mu]}$. Each particular copy of $\mrep{[\mu]}$ is an energy eigenspace and there may be multiple copies of $\mrep{[\mu]}$ corresponding to the same energy.

For one-component fermions and bosons, the spin Hilbert space $\SHS$ is trivial and the total Hilbert space is just $\HS\sim\KHS$. Only states in the totally symmetric subspace $\KHS^{[N]}$ can be populated by one-component bosons, and only the totally antisymmetric subspace $\KHS^{[1^N]}$ is available for one-component fermions.
For fermions and bosons with $J>1$ components, the spin Hilbert space $\SHS\sim \mathbb{C}^{J^N}$ can `carry' some of the symmetry or antisymmetry required for bosons or fermions and so the mixed symmetry subspaces of $\KHS$ are relevant for identical bosons and fermions.

One way to treat symmetrization of identical particles with internal components that do not participate in the Hamiltonian is to reduce $\SHS$ into $\mathrm{S}_N$ irrep spaces
\begin{equation}\label{eq:khsirreps}
\SHS = \bigoplus_{[\mu]\in P(N)} \SHS^{[\mu]}
\end{equation}
using standard techniques (c.f.\ \cite{Hamermesh, Chen, Jakubczyk}) and then reduce the tensor product $\HS= \KHS \otimes \SHS $ into irreps using the Clebsch-Gordan series for $\mathrm{S}_N$. 
When the internal components are spin, then $J=2s+1$ and $\SHS$ also carries a reducible representation of $\mathrm{SU}(2)$. The total spin operator $\hat{S}^2 = \sum_i \hat{S}_i^2$ and total spin component $\hat{S}_z = \sum_i \hat{S}_{z,i}$ are invariant under $\mathrm{S}_N$ and can be diagonalized along with the $\mathrm{S}_N$ irreps. 

Explaining this topic exceeds the ambitions of the present article, but two important results are:
\begin{itemize}
\item For each irrep $[\nu]$ in the decomposition of $\SHS$, there will be a single bosonic state for each copy of $\mrep{[\nu]}$ in the  sector $\KHS^{[\nu]}$.
\item For each irrep $[\nu]$ in the decomposition of $\SHS$  there is  a single fermionic state for each copy of $\mrep{[\nu]^\top}$ in the in the sector $\KHS_{[\nu]^\top}$.
\end{itemize}
Examples with two and three particles were provided in the proceeding article. As an example for four particles, consider the case of spin-$1/2$ fermions. Two-component particles can only have internal states with at most two-row $\mathrm{S}_4$ irreps, and the spin Hilbert space can be reduced in several ways:
\begin{equation}
\SHS = \SHS^{[4]} \oplus \SHS^{[31]} \oplus \SHS^{[2^2]}
\end{equation}
where 
\begin{eqnarray}
\SHS^{[4]} &=& \SHS_{\tiny \young(\uparrow\uparrow\uparrow\uparrow)} \oplus \SHS_{\tiny \young(\uparrow\uparrow\uparrow\downarrow)} \oplus \SHS_{\tiny \young(\uparrow\uparrow\downarrow\downarrow)} \oplus \SHS_{\tiny \young(\uparrow\downarrow\downarrow\downarrow)} \oplus \SHS_{\tiny \young(\downarrow\downarrow\downarrow\downarrow)}\nonumber\\
&\sim& 5 \mrep{[4]} \sim D^{(s=2)} \nonumber\\
\SHS^{[31]} &=& \SHS_{\tiny \young(\uparrow\uparrow\uparrow,\downarrow)} \oplus \SHS_{\tiny \young(\uparrow\uparrow\downarrow,\downarrow)} \oplus \SHS_{\tiny \young(\uparrow\downarrow\downarrow,\downarrow)} \nonumber\\
&\sim& 3 \mrep{[31]} \sim 3 D^{(s=1)} \nonumber\\
\SHS^{[2^2]} &=&  \SHS_{\tiny \young(\uparrow\uparrow,\downarrow\downarrow)} \sim \mrep{[2^2]} \sim 2 D^{(s=0)}
\end{eqnarray}
where $D^{(s)}$ denotes $\mathrm{SU}(2)$ irrep spaces. This reduction can be used to find the degeneracy and spin possible for spatial states: for every level in $\KHS^{[1^4]}$, there are the five states with total spin $2$; for every level in $\KHS^{[21^2]}$, there are nine spin-$1$ states, three which each possible $z$-component; and for every level in $\KHS^{[2^2]}$, there will be two spin-$0$ states. Constructing the explicit spin-spatial states in terms of the particle basis is a technical challenge that increases with complexity as $N$ and $J$ get bigger, but it is an algebraically solvable problem. See \cite{Cui2014, Yurovsky2014, Yurovsky2015} for recent applications of these methods to trapped particles with spin.

In summary, reducing the spatial Hilbert space into irreps of $\mathrm{S}_N$ is useful for symmetrizing identical particles, as well as understanding the degeneracy of energy eigenstates and how the energy levels split and combine as the trap and interaction are changed. The rest of this article shows how additional symmetries of the interaction and the trap enrich this structure.

\section{Non-Interacting Particles}

For the non-interacting $N$-particle system, denote the configuration space symmetry group as $\mathrm{C}_N^0$ and the kinematic symmetry group as $\mathrm{K}_N^0$. The total non-interacting system inherits a minimal configuration space symmetry group and a minimal kinetic symmetry group  from its construction out of one-particle systems:
\begin{equation}\label{eq:mincon}
\mathrm{C}_N^0 \supseteq  \mathrm{P}_N \ltimes \mathrm{C}_1^{\times N}\ \mbox{and}\ 
\mathrm{K}_N^0 \supseteq  \mathrm{P}_N \ltimes \mathrm{K}_1^{\times N},
\end{equation}
where $\mathrm{G}^{\times N}$ means the group constructed from $N$-fold direct product of $\mathrm{G}$ with itself and the particle permutation group $\mathrm{P}_N$ acts via a semidirect product $\ltimes$ on the abelian, normal subgroups of $\mathrm{C}_1^{\times N}$ or $\mathrm{K}_1^{\times N}$ by rearranging terms in the direct product.  

Before diving into representation theory, let us physically motivate this construction.
In the case of a symmetric well, each particle's individual parity operator $\hat{\Pi}_i$ commutes with all the other parity operators and with the total Hamiltonian $\hat{H}_0^N$. 
Therefore the configuration space symmetry $\mathrm{C}_N^0$ must at least have an abelian subgroup $\mathrm{O}(1)^{\times N}$ generated by the $N$ commuting, parity operators. Similarly, each individual particle's Hamiltonian $\hat{H}^1_i$ commutes with the total Hamiltonian. This implies that each particle's individual time evolution is still a good symmetry: the clocks of individual particles are not synchronized unless there are interactions\footnote{This observation may seem obvious, but it was a remark to this effect in \cite{Fernandez} discussing the similarly separable problem of a particle in a cubic box that led me to understand the connection between permutation modules and $\mathrm{K}_N^0$.}. Therefore the kinematic group  $\mathrm{K}_N^0$ must at least have an abelian subgroup $\mathrm{T}_t^{\times N}$ generated by each individual particle's time evolution. Because particle exchanges commute with the total Hamiltonian,  $\mathrm{P}_N \sim \mathrm{S}_N$ must also be subgroup of both $\mathrm{C}_N^0$ and $\mathrm{K}_N^0$. However particle exchanges do not commute with the single-particle parities and Hamiltonians, e.g.\ $(ij)\Pi_i = \Pi_j (ij)$. The semidirect product in (\ref{eq:mincon}) captures that structure of the minimal subgroups of $\mathrm{C}_N^0$ and $\mathrm{K}_N^0$ the same way the semidirect product is useful for describing groups of affine transformations as semidirect product of linear transformations and translations.

In the subsections below, the groups $\mathrm{C}_N^0$ and $\mathrm{K}_N^0$ and their irreps are explored in more detail. See Table \ref{tab:lowNgroups} for a summary of results for $N=2$, $3$ and $4$.

\begin{table}[t]
\caption{This table provides information about the kinematic symmetry group $\mathrm{K}_N^0$ and configuration space symmetry group $\mathrm{C}_N^0$ for $N=2$, $3$ and $4$ identical non-interacting particles in asymmetric, symmetry and harmonic traps. See text to understand notation for structure of $\mathrm{C}_N^0$ and $\mathrm{K}_N^0$. For each group, the dimensions of the unitary irreducible representations (irreps) is provided, and the number of inequivalent irreps with that dimension is indicated by the superscript. The symbol $\lambda\in\mathbb{N}$ labels inequivalent $\mathrm{O}(N)$ irreps and $X\in\mathbb{N}$ labels inequivalent $\mathrm{U}(N)$ irreps. Note that for $N=2$, there is an infinite tower of inequivalent irreps of $\mathrm{O}(2)$ for $\lambda>0$. The Coxeter notation and order is given for the finite-order configuration space groups, and the Sch\"onflies point group notation is also given for $N=2$ and $3$.}
\centering
\label{tab:lowNgroups}
\begin{tabular}{|c|c|c|c|}
\hline
 & Asym. Trap &  Sym. Trap &  Harm. Trap\\
\hline\hline\hline
$\mathrm{C}_2^0$ & $\mathrm{S}_2$ & $\mathrm{S}_2 \!\ltimes\! \mathrm{O}(1)^{\times 2}$ & $\mathrm{O}(2)$ \\
\hline
Irreps & $1^2$ & $1^4,2$ & $1,2$ \\
\hline
Point & $\mathrm{D}_1$ & $\mathrm{D}_4$ & $\mathrm{O}(2)$  \\
\hline
Coxeter & $\mathrm{A}_1\sim [\,]$ & $\mathrm{BC}_2 \sim [4]$ &  ---\\
\hline
 Order & $2$ & $8$ & $\infty$ \\
\hline\hline
$\mathrm{K}_2^0$  & $\mathrm{S}_2 \!\ltimes\! \mathrm{T}_t^{\times 2}$ & $\mathrm{S}_2 \!\ltimes\! (\mathrm{O}(1) \!\times\! \mathrm{T}_t)^{\times 2}$ & $\mathrm{U}(2)$ \\
\hline
Irreps & $1,2$ & $1^2,2^3$ & $(X+1)$\\
\hline\hline\hline
$\mathrm{C}_3^0$ & $\mathrm{S}_3$ & $\mathrm{S}_3 \!\ltimes\! \mathrm{O}(1)^{\times 3}$ & $\mathrm{O}(3)$ \\
\hline
Irreps & $1^2,2$ & $1^4,2^2,3^4$ & $2 \lambda +1$\\
\hline
Point & $\mathrm{C}_{3v}$ & $\mathrm{O}_h$ & $\mathrm{O}(3)$ \\
\hline
Coxeter  & $\mathrm{A}_2\sim [3]$ & $\mathrm{BC}_3 \sim [4,3]$ & --- \\
\hline
Order & $6$ & $48$ & $\infty$ \\
\hline
\hline
$\mathrm{K}_3^0$ & $\mathrm{S}_3 \!\ltimes\! \mathrm{T}_t^{\times 3} $ & $\mathrm{S}_3 \!\ltimes\! (\mathrm{O}(1)\! \times\! \mathrm{T}_t)^{\times 3}$ & $\mathrm{U}(3)$ \\
\hline
Irreps & $1,3,6$ & $1^2,3^4,6^4$ & $\frac{(X+2)!}{2 X!}$ \\
\hline
\hline\hline
$\mathrm{C}_4^0$  & $\mathrm{S}_4$ & $\mathrm{S}_4 \!\ltimes\! \mathrm{O}(1)^{\times 4}$ & $\mathrm{O}(4)$ \\
\hline
Irreps & $1^2,2,3^2$ & $1^4,2^2,3^4,4^4,6^4,8^2$ & $(\lambda +1)^2$ \\
\hline
Coxeter  & $\mathrm{A}_3\sim [3,3]$ & $\mathrm{BC}_4 \sim [4,3,3]$ & --- \\
\hline
Order & $24$ & $384$ & $\infty$  \\
\hline
\hline
$\mathrm{K}_4^0$ & $\mathrm{S}_4 \!\ltimes\! \mathrm{T}_t^{\times 4} $ & $\mathrm{S}_4 \!\ltimes\! (\mathrm{O}(1)\! \times\! \mathrm{T}_t)^{\times 4}$ & $\mathrm{U}(4)$ \\
\hline
Irreps & $1,4,6,12,24$ & $1^2,4^4,6^3,12^6,24^5$ & $\frac{(X+3)!}{6 X!}$  \\
\hline

\end{tabular}

\end{table}

\subsection{Configuration Space Symmetry Group $\mathrm{C}_N^0$}

The configuration space $\mathcal{Q}^N = \mathbb{R}^N$ of $N$ particles in one dimension is isomorphic to one particle in $N$ dimensions, and therefore low $N$ situations can be visualized and described using the terms and techniques of familiar geometry. Additionally, for asymmetric and for most symmetric wells, the configuration space symmetry group is a finite-order point group. Finite-order point groups are subgroups of the orthogonal transformations $\mathrm{O}(N)$ on $\mathcal{Q}^N$. Point groups in all dimensions are completely characterized and classified.  Point groups in two and three dimensions are familiar to many physicists from applications in chemical and solid state physics and Sch\"onflies notation is standard. There are several notations for extensions of these groups to higher dimensions, but Coxeter notation~\cite{Coxeter} is most convenient for my purposes because of its connection to the theory of Weyl groups and Lie algebras, which finds application in the closely related theory of the Bethe ansatz solutions~\cite{Gaudin1971, Gaudin}.

For the asymmetric well, the one-particle configuration symmetry group is trivial $\mathrm{C}_1 \sim \mathrm{Z}_1$ and the configuration space symmetry $\mathrm{C}_N^0$ is isomorphic to the permutation group $\mathrm{S}_N$. Each particle permutation is realized by a geometrical transformation of configuration space $\mathcal{Q}^N$, for example:
\begin{itemize}
\item Two-cycles $(ij)$ are reflections across the $(N\!-\!1)$-dimensional hyperplane $\mathcal{V}_{ij}\subset \mathcal{Q}^N$ defined by $q_i = q_j$.
\item Three-cycles $(ijk)$ are generated by two overlapping two-cycles $(ij)(jk)$. They are realized by simple rotations by $\pm 2\pi/3$ in the plane perpendicular to the $(N\!-\!2)$-dimensional hyperplane $\mathcal{V}_{ij}\cap \mathcal{V}_{jk}$.
\item Double two-cycles $(ij)(kl)$ are generated by two non-overlapping two-cycles $(ij)(kl)$. These are double reflections across orthogonal hyperplanes $\mathcal{V}_{ij}$ and $\mathcal{V}_{kl}$, and they are equivalent to a simple rotation by $\pm \pi$ in the plane perpendicular to the $(N\!-\!2)$-dimensional hyperplane $\mathcal{V}_{ij}\cap\mathcal{V}_{kl}$.
\item Four-cycles $(ijkl)$ are simple rotoreflections (or improper reflections). They are realized by a rotation by $\pm \pi/2$ in the plane perpendicular to $\mathcal{V}_{ik}\cap\mathcal{V}_{jl}$, followed by reflection across the same plane.
\end{itemize}
Longer cycle structures correspond to elements of equivalence classes of higher-dimensional orthogonal transformations, such as compound reflections, compound rotations and compound rotoreflections. One way to derive these properties is to construct the defining representation of $\mathrm{S}_N$ on the set of basis vectors $\{\hat{q}_1, \hat{q}_2, \ldots, \hat{q}_N\}$. For example, the matrix representing the element $(12345)=\{23451\}\in\mathrm{S}_5$ is
\begin{equation}
\underline{O}(12345) = \left( \begin{array}{ccccc} 0 & 1 & 0 & 0 & 0 \\ 0 & 0 & 1 & 0 & 0 \\ 0 & 0 & 0 & 1 & 0 \\ 0 & 0 & 0 & 0 & 1 \\ 1 & 0 & 0 & 0 & 0 \end{array} \right)
\end{equation}
The eigenvalues of $\underline{O}(12345)$ are $\{ 1, \exp(\pm 4\pi i/5), \exp(\pm 2\pi i/5)\}$, corresponding to a compound double rotation of $4 \pi/5$ and $2\pi/5$ in two orthogonal planes. The eigenvalue $1$ corresponds to the eigenvector ${\bf q} = (1,1,1,1,1)/\sqrt{5}$ which remains invariant under all permutations of five particles.

The geometrical realization of $\mathrm{P}_N$ is equivalent to the point symmetry group of a regular $N$-simplex: the digon for $N=2$, the triangle for $N=3$, the tetrahedron for $N=4$, the pentachoron for $N=5$, etc. These are the finite Coxeter reflection groups $\mathrm{A}_{N-1}$. Note also that conjugacy classes of $\mathrm{P}_N$ are associated to equivalence classes of $\mathrm{O}(N\!-\!1)$, e.g.\  elements of $\mathrm{P}_4 \sim \mathrm{S}_4$ are reflections, simple rotations, and simple rotoreflections, and these are the three equivalence classes of $\mathrm{O}(3)$. This restriction to equivalence classes of  $\mathrm{O}(N\!-\!1)$  is because the one-dimensional manifold defined by $q_1=q_2 = \cdots = q_N$ is invariant under the geometric realization of every $p\in\mathrm{P}_N$.

For a symmetric well, the configuration space symmetry $\mathrm{C}_N^0$ must contain $\mathrm{S}_N \ltimes \mathrm{O}(1)^{\times N}$, the semidirect product of the symmetric group $\mathrm{S}_N$ on the $N$-fold tensor product of the reflection group $\mathrm{O}(1)$. Each individual particle parity operator $\Pi_i$ is realized by reflections across the $(N\!-\!1)$-dimensional hyperplane given by $q_i=0$.  The total parity $\Pi$ is the product of all the individual parities
\begin{equation}
\Pi = \Pi_1\Pi_2\cdots \Pi_N.
\end{equation}
The group $\mathrm{S}_N \ltimes \mathrm{O}(1)^{\times N}$ has order $2^N N!$ and it is also known as the hyperoctahedral group~\cite{Frame, Goodman}. These are the finite Coxeter reflection groups labeled $\mathrm{BC}_{N}$. The $N=2$ and $N=3$ dimensional examples are the point groups of a square and cube, respectively, and their irreps (and the reduction of those irreps by the subgroup $\mathrm{S}_N$) are discussed in the previous article and summarized in Table~\ref{tab:lowNgroups}. The higher-dimensional point groups for the hypercubes are less well-known in physics but their irreps can be found by induction from the normal subgroup $\mathrm{O}(1)^{\times N}$. That is how the irrep dimension and multiplicity is calculated for $N=4$ in Table~\ref{tab:lowNgroups}.

For a harmonic well, there is an emergent multi-particle symmetry and $\mathrm{C}_N^0$ is larger than minimal symmetry inherited from the construction (\ref{eq:mincon}). There is full rotational and reflectional symmetry in $\mathcal{Q}^N$ and so $\mathrm{C}_N^0\sim \mathrm{O}(N)$ is the full orthogonal group in $N$ dimensions. The hyperspherical representation of this group is well-known: irreps are labeled by $\lambda$ and have the dimension~\cite{Avery}
\begin{equation}
d(\mathrm{O}(N);\lambda) = \frac{(N + 2 \lambda - 2) (N + \lambda - 3)!}{\lambda! (N-2)!}
\end{equation}
for $N>2$. This formula gives the familiar results $d(\mathrm{O}(3);\lambda)=2\lambda+1$ and $d(\mathrm{O}(4);\lambda)= (\lambda+1)^2$. This case has been examined in more detail in \cite{Harshman2014} and further applications are in preparation.

Note that if $\mathrm{C}_N^0$ were the only symmetry of the non-interacting Hamiltonian $\hat{H}_0^N$, then we would expect that the energy levels would have degeneracies corresponding to the dimensions of those irreps. However, they certainly do not and the explanation of spectral degeneracies requires the consideration of the kinematic symmetry group $\mathrm{K}_N^0$.

\subsection{Kinematic Symmetry Group $\mathrm{K}_N^0$}

Denote the non-interacting $N$-particle spectrum by $\sigma_N^0 =\{E_0, E_1, E_2, \cdots\}$, which is still discrete but no longer non-degenerate. The spatial Hilbert space is decomposable into energy eigenspaces
\begin{equation}\label{eq:energydecomp}
\KHS = \bigoplus_k \KHS^{k}
\end{equation}
where $\hat{H}^N_0 \KHS^{k} = E_k \KHS^{k}$. If the kinematic symmetry group $\mathrm{K}_N^0$ has been completely and correctly identified, then each energy eigenspace $\KHS^k$ will carry an irreducible representation of $\mathrm{K}_N^0$. In that case, the dimensions of the irreps of $\mathrm{K}_N^0$ correspond to the degeneracies of energies $E_k\in\sigma_N^0$. This section demonstrates that unless there are accidental degeneracies or emergent few-body symmetries, then the minimal kinematic group $\mathrm{P}_N \ltimes \mathrm{K}_1^{\times N}$ is sufficient to explain the degeneracies of $E_k \in \sigma_N^0$. The irreps of $\mathrm{P}_N \ltimes \mathrm{K}_1^{\times N}$ are isomorphic to the permutation modules described in Section 2.2 and their reduction into $\mathrm{S}_N$ irreps is algebraically solvable in terms of one-particle observables and particle permutation operators.

The elements of $\sigma_N^0$ and their degeneracies can be determined by forming compositions of one-particle energies $\epsilon_n$ in the single particle spectrum $\sigma_1$. The energy level with  composition $\pp{\nu}=\pp{0^{\nu_0} 1^{\nu_1} \ldots}$ has energy $E_\pp{\nu} = \nu_0 \epsilon_0 + \nu_1 \epsilon_1 + \cdots$. Only a partial ordering of $\sigma_N^0$ is possible unless the specific one-particle energies $\epsilon_n\in\sigma_1$ are known; see previous article for examples.

The tensor product of $N$ non-interacting basis states is compactly denoted by
\begin{equation}
\kt{\bf n} \equiv \kt{n_1} \otimes  \kt{n_2} \otimes \cdots \otimes \kt{n_N},
\end{equation}
or alternatively $\kt{\alpha\beta\cdots}$. Call this basis, where each particle has a definite state, the ``particle basis''. Note that the particle basis wave functions
\begin{equation}
\Phi_{\bf n}({\bf q}) = \bk{\bf q}{\bf n} = \prod_{i=1}^N \phi_{n_i}(q_i)
\end{equation}
can always be chosen as real functions for the trapped system.
Each $N$-particle tensor product basis vector $\kt{{\bf n}}$ is an eigenvector of $\hat{H}^0_N$ with energy $E_\pp{\nu}$ given by the composition $\pp{\nu}$ of ${\bf n}$. The degeneracy $d\pp{\nu}$ of the energy level $E_\pp{\nu}$ is the number of particle basis vectors with that composition. As explained in the previous section, this degeneracy is determined by the shape $[\nu]$ of the composition $\pp{\nu}$. 

The spatial Hilbert space is also decomposable into subspaces $\mathcal{K}^\pp{\nu}$ spanned by particle basis vectors with composition $\pp{\nu}$
\begin{equation}\label{eq:compdecomp}
\KHS = \bigoplus_\pp{\nu} \KHS^\pp{\nu}.
\end{equation}
Note that this decomposition is not the same as (\ref{eq:khsirreps}). The sectors $\KHS^{[\mu]}$ are infinite towers of $\mathrm{S}_N$ irrep spaces $\mrep{[\mu]}$, each with dimension $d[\mu]$, whereas $\KHS^\pp{\nu}$ are the composition subspaces with dimension $d\pp{\nu}$. Each $\mathcal{K}^\pp{\nu}$ is an irrep of $\mathrm{S}_N \ltimes \mathrm{T}_t^{\times N}$ and isomorphic to a permutation module $M^{[\nu]}$. This is demonstrated in the next subsection by explicit construction of the $d\pp{\nu}$-dimensional irrep  of  $\mathrm{S}_N \ltimes \mathrm{T}_t^{\times N}$.

\subsubsection{Irreps of $\mathrm{S}_N \ltimes \mathrm{K}_1^{\times N}$}\label{sec:induced}

An element $(p,{\bf t})$ of $\mathrm{S}_N \ltimes \mathrm{T}_t^{\times N}$ is a pair formed by a permutation $p\in\mathrm{S}_N$ and a real $N$-tuple ${\bf t}=\langle t_1,t_2,\ldots,t_N \rangle \in\mathrm{T}_t^{\times N}$.  The group multiplication rule in $\mathrm{S}_N \ltimes \mathrm{T}_t^{\times N}$ is
\begin{equation}\label{eq:K0Nrule}
( p';{\bf t}')\cdot( p;{\bf t})= (p'p; {\bf t} + \underline{O}^\top(p){\bf t}')
\end{equation}
so that
\begin{equation}
(p, {\bf t}) = (p, {\bf 0})\cdot (e, {\bf t}),
\end{equation}
where $e$ is the identity in $\mathrm{S}_N$  and ${\bf 0}$ is the identity in $\mathrm{T}_t^{\times N}$. 
The $N \times N$ matrix $\underline{O}(p)$ permutes the components of an $N$-dimensional vector ${\bf x}$:
\begin{eqnarray}
\underline{O}(p) {\bf x} &=& \underline{O}(p) \langle x_1, x_2, \ldots, x_N \rangle\nonumber\\
&=& \langle x_{p_1}, x_{p_2}, \ldots, x_{p_N} \rangle.
\end{eqnarray}
The matrix $\underline{O}^\top(p) = \underline{O}^{-1}(p) = \underline{O}(p^{-1})$ is orthogonal and all matrix elements are zero except a single one in each row and column\footnote{The representation $\underline{O}$ is another example of the defining representation of $\mathrm{S}_N$. By a similarity transformation $\underline{J}^\top \underline{O} \, \underline{J}$ it can be decomposed into $\mathrm{S}_N$ irreps $[N]\oplus[(N\!-\!1)\,1]$. The set of similarity transformations $\underline{J}$ that reduce the defining representation is the equivalence class of normalized Jacobi coordinate systems for $N$ one-dimensional particles.}.

The choice of the transpose matrix as the automorphism on $\mathrm{T}_t^{\times N}$ in the multiplication rule (\ref{eq:K0Nrule}) is so  that the unitary representation $\hat{U}(p, {\bf t})$ of $\mathrm{S}_N \ltimes \mathrm{T}_t^{\times N}$ has a natural realization on the particle basis $\kt{\bf n}$:
\begin{eqnarray}\label{eq:K0Nrep}
\hat{U}(p,{\bf 0}) \kt{n_1\,n_2,\ldots,n_N} &=& \kt{n_{p_1},n_{p_2},\ldots,n_{p_N}} = \kt{\underline{O}(p) {\bf n}} \nonumber\\
\hat{U}(e,{\bf t}) \kt{{\bf n}} &=& \exp(-i \bep \cdot {\bf t})  \kt{{\bf n}}\nonumber\\
\hat{U}(p,{\bf t}) \kt{{\bf n}} &=& \exp(-i \bep \cdot {\bf t})  \kt{\underline{O}(p) {\bf n}},
\end{eqnarray}
where $\bep = \{\epsilon_{n_1}, \epsilon_{n_2},\ldots,\epsilon_{n_N}\}$.

The orbit of any $N$ particle basis vector $\kt{\bf n}$ with composition $\pp{\nu}$ under the representation (\ref{eq:K0Nrep}) spans the composition space $\mathcal{K}^\pp{\nu}$. The unitary matrix representation $\underline{D}^\pp{\nu}(p,{\bf t})$ on $\mathcal{K}^\pp{\nu}$ defined by
\begin{equation}
\hat{U}(p, {\bf t}) \kt{\bf n} = \sum_{{\bf n}' \in ({\bf n})} \underline{D}^\pp{\nu}_{{\bf n}'{\bf n}}(p,{\bf t})\kt{{\bf n}'}
\end{equation}
is $d\pp{\nu}$ dimensional and irreducible with respect to $\mathrm{S}_N \ltimes \mathrm{T}_t^{\times N}$. The action of the subgroup $\underline{D}^\pp{\nu}(p,{\bf 0})$ on $\mathcal{K}^\pp{\nu}$ is unitarily equivalent to the action of the symmetric group on permutation module $M^{[\nu]}$. Therefore, irreps of $\mathrm{S}_N \ltimes \mathrm{T}_t^{\times N}$ fall into equivalence classes, one for each shape $[\nu]\in P(N)$.

An alternate derivation of the irreps of $\mathrm{S}_N \ltimes \mathrm{T}_t^{\times N}$ uses the technique of induced representations: the character of the normal, abelian subgroup  $\mathrm{T}_t^{\times N}$ is the exponential $\exp(-i \bep \cdot {\bf t})$ and it is determined by the sequence of energies $\bep$, or equivalently by the sequence ${\bf n}$. The distinct orbits of ${\bf n}$ under $\mathrm{S}_N$ are classified by the shape of the composition $[\nu]$. The `little group' of a composition $\pp{\nu}$, i.e.\ the group of transformations that leaves a canonical representative $\tilde{\bf n}\in\pp{\nu}$ invariant, is $\mathrm{S}_\pp{\nu} = \mathrm{S}_{\nu_1} \times \mathrm{S}_{\nu_2}\times \cdots \times \mathrm{S}_{\nu_r}$. The degeneracy $d\pp{\nu}$ is the order of the coset $\mathrm{S}_N/\mathrm{S}_\pp{\nu}$ and there is a basis vector of the irrep for each element of the coset. For example, when $\pp{\nu}=\pp{012^24}$,  choose the normal-ordered representative sequence $\tilde{\bf n}= \langle 0,1,2,2,4 \rangle$. The partition $[\nu] = [21^3]$ has little group $\mathrm{S}_{[21^3]}\sim\mathrm{S}_2$ with two elements $e$ and $(34)$, corresponding to exchange of the identical symbols. The construction or irreps by induced representations guarantees irreducibility. Note that 
$\mathrm{S}_\pp{\nu}$ is not the same thing as the symbol permutation group, which would be $\mathfrak{S}_{\pp{012^24}}\sim\mathrm{S}_3$.

\subsubsection{The Double Tableaux Basis for Eigenstates of $\hat{H}_0^N$}

The previous subsection establishes that composition spaces $\KHS^\pp{\nu}$ are irreps of $\mathrm{S}_N \ltimes \mathrm{T}_t^{\times N}$ and isomorphic to permutation modules $M^{[\nu]}$. The particle basis $\kt{\bf n}$ provides a natural basis for this group that diagonalizes the time translation subgroup $\mathrm{T}_t^{\times N}$. However, for several practical reasons, the reduction of $\KHS^\pp{\nu}$ by the $\mathrm{S}_N$ subgroup is often useful:
\begin{itemize}
\item This reduction allows the use of the double tableau basis $\kt{W\,Y}$. The irreps that occur in $\KHS^\pp{\nu}$ are labeled by Weyl tableaux $W$ and the shapes of the tableaux $[W]$ indicate the $\mathrm{S}_N$ irrep $\mathcal{M}^{[W]}$. The specific Young tableau $Y\in[W]$ labels the irrep basis. See Ref.\ \cite{Chen} for  a detailed derivation of this for any $N$ and explicit construction of the complete set of commuting observables for this basis out of conjugacy class operators for canonical and non-canonical subgroup chains. The previous article also includes an extended example with $N=3$. 
\item For spinless particles, the reduction of $\KHS^\pp{\nu}$ into irreps of $\mathrm{S}_N$ provides a method for handling identical particle symmetrization. For particles with spin, the reduction of the spatial Hilbert space into $\mathrm{S}_N$ sectors provides a methods for calculating degeneracies, observables, and basis vectors, c.f.\ Section 2.3.
\item When interactions are added to the Hamiltonian, the kinematic symmetry $\mathrm{S}_N \ltimes \mathrm{T}_t^{\times N}$ will be broken, but the $\mathrm{S}_N$ symmetry will remain. One consequence is that (unless there are accidental or emergent symmetries) the degenerate energy levels of each composition space $\KHS^\pp{\nu}$ will split into levels, specifically, one $d[\mu]$-degenerate level for each copy of $\mrep{[\mu]}$ in $\KHS^\pp{\nu} \sim M^{[\nu]}$.
\item Also, when interactions are incorporated there are only matrix elements between non-interacting basis vectors within the same $\mathrm{S}_N$ irrep. Therefore, exact diagonalization in the double tableau basis requires fewer matrix elements to achieve the same accuracy, and the energy spectra are more easily interpreted in terms of spectroscopy and selection rules, c.f.\ Section 4.4.
\end{itemize}
 
 As an example of the double tableaux basis with $N=4$, consider the reduction of a composition space from each equivalence class of $\mathrm{S}_4 \ltimes \mathrm{T}_t^{\times 4}$ irreps:
\begin{eqnarray}
\KHS^\pp{\alpha^4} &=& \KHS^{\tiny \young(\alpha\alpha\alpha\alpha)} \nonumber\\
\KHS^\pp{\alpha^3\beta} &=& \KHS^{\tiny \young(\alpha\alpha\alpha\beta)} \oplus \KHS^{\tiny \young(\alpha\alpha\alpha,\beta) }\nonumber\\
\KHS^\pp{\alpha^2\beta^2} &=& \KHS^{\tiny \young(\alpha\alpha\beta\beta)} \oplus \KHS^{\tiny \young(\alpha\alpha\beta,\beta)} \oplus \KHS^{\tiny \young(\alpha\alpha,\beta\beta)}\nonumber\\
\KHS^\pp{\alpha^2\beta\gamma} &=& \KHS^{\tiny \young(\alpha\alpha\beta\gamma)} \oplus \KHS^{\tiny \young(\alpha\alpha\beta,\gamma)} \oplus \KHS^{\tiny \young(\alpha\alpha\gamma,\beta)} \oplus \KHS^{\tiny \young(\alpha\alpha,\beta\gamma)} \oplus \KHS^{\tiny \young(\alpha\alpha,\beta,\gamma)}\nonumber\\
\KHS^\pp{\alpha\beta\gamma\delta} &=& \KHS^{\tiny \young(\alpha\beta\gamma\delta)} \oplus \KHS^{\tiny \young(\alpha\beta\gamma,\delta)} \oplus \KHS^{\tiny \young(\alpha\beta\delta,\gamma)} \oplus \KHS^{\tiny \young(\alpha\gamma\delta,\beta)} \oplus \KHS^{\tiny \young(\alpha\beta,\gamma\delta)} \oplus \KHS^{\tiny \young(\alpha\gamma,\beta\delta)} \nonumber\\
&& {} \oplus \KHS^{\tiny \young(\alpha\beta,\gamma,\delta)} \oplus \KHS^{\tiny \young(\alpha\gamma,\beta,\delta)} \oplus \KHS^{\tiny \young(\alpha\delta,\beta,\gamma)} \oplus \KHS^{\tiny \young(\alpha,\beta,\gamma,\delta)}
\end{eqnarray}
Each composition space $\KHS^\pp{\nu}$ is isomorphic to the permutation module $M^{[\nu]}$ and each subspace $\KHS^W$ is isomorphic to the $\mathrm{S}_4$ irrep $\mathcal{M}^{[W]}$.

Depending on the shape of the composition, the composition space $\KHS^\pp{\nu}$ also carries a representation of another group, called the state permutation group $\mathfrak{P}_\pp{\nu}$. First consider an example with $\KHS^\pp{\alpha^2\beta\gamma}$. This space is $d\pp{\nu}=12$ dimensional and has basis vectors (here in the particle basis):
\begin{eqnarray}\label{bas:211}
\kt{\alpha\alpha\beta\gamma}, \kt{\alpha\alpha\gamma\beta}, \kt{\alpha\beta\alpha\gamma}, \kt{\alpha\beta\gamma\alpha}, \kt{\alpha\gamma\alpha\beta}, \kt{\alpha\gamma\beta\alpha},\nonumber\\ \kt{\beta\alpha\alpha\gamma}, \kt{\beta\alpha\gamma\alpha}, \kt{\beta\gamma\alpha\alpha}, \kt{\gamma\alpha\alpha\beta}, \kt{\gamma\alpha\beta\alpha}, \kt{\gamma\beta\alpha\alpha}.
\end{eqnarray}
The action of the particle permutation group $\mathrm{S}_4$ on this basis is given by (\ref{eq:K0Nrep}). However, there is another symmetry that leaves the space $\KHS^\pp{\alpha^2\beta\gamma}$ invariant: switching $\beta$ and $\gamma$. Denote the operator that switches these states as $\hat{U}(\beta\gamma)$. This operator shuffles the basis vectors (\ref{bas:211}) into
\begin{eqnarray}
\kt{\alpha\alpha\gamma\beta}, \kt{\alpha\alpha\beta\gamma}, \kt{\alpha\gamma\alpha\beta}, \kt{\alpha\gamma\beta\alpha}, \kt{\alpha\beta\alpha\gamma},  \kt{\alpha\beta\gamma\alpha},\nonumber\\ \kt{\gamma\alpha\alpha\beta}, \kt{\gamma\alpha\beta\alpha}, \kt{\gamma\beta\alpha\alpha}, \kt{\beta\alpha\alpha\gamma}, \kt{\beta\alpha\gamma\alpha}, \kt{\beta\gamma\alpha\alpha}.
\end{eqnarray}
The operator $\hat{U}(\beta\gamma)$ generates a group isomorphic to $\mathrm{S}_2$ but distinct from any $\mathrm{S}_2$ subgroup of the particle permutation group $\mathrm{P}_4$. Call this group $\mathfrak{P}_\pp{\alpha^2\beta\gamma}$. Every irrep of $\mathrm{S}_4 \ltimes \mathrm{T}_t^{\times 4}$ with shape $[21^2]$ carries a twelve-dimensional reducible representation of $\mathfrak{P}_\pp{\alpha^2\beta\gamma}$. Note that  the $\mathfrak{P}_\pp{\alpha^2\beta\gamma}$ symmetry is not a symmetry of the full Hamiltonian $\hat{H}_N^0$ and most composition subspaces are not invariant under $\hat{U}(\beta\gamma)$. However, it is useful for analyzing specific composition spaces like $\KHS^\pp{\alpha^2\beta\gamma}$. Another way to say this is that $\hat{U}(\beta\gamma)$ and $\hat{H}^0_N$ commute when restricted to the spaces like $\KHS^\pp{\alpha^2\beta\gamma}$, even though the operators defined on the entire spatial Hilbert space $\KHS$ do not.

Referring back to Section 2.2, we see that particle permutation symmetry $\mathrm{P}_N$ places the role of sequence permutation symmetry and state permutation symmetry $\mathfrak{P}_\pp{\nu}$ plays the role of symbol permutation symmetry. The representation of $\mathrm{S}_N \ltimes \mathrm{T}_t^{\times N}$ is irreducible on a composition space $\KHS^\pp{\nu}\sim M^{[\nu]}$. The operators $\hat{U}(p)$  representing the particle permutation $p\in\mathrm{P}_N$ mix double tableau basis vector  $\kt{W\,Y}$  with the same $W$ and different $Y$ and the operators $\hat{U}(\mathfrak{p})$  representing the particle permutation $\mathfrak{p}\in\mathfrak{P}_\pp{\nu}$ mix basis vectors with the same $Y$ and different $W$.

The coefficients $\bk{\bf n}{W\,Y}$ that connect the double tableau basis $\kt{W\,Y}$ and the particle basis $\kt{\bf n}$ can be found using a variety of methods. In the previous article, I follow the conventions and methods of Ref.~\cite{Chen} for two and three particles. These methods can be extended to more particles, and although it becomes more complicated as $N$ grows larger, it is still algebraically solvable for any $N$. Explicit values for $\bk{\bf n}{W\,Y}$ are not necessary for the results that follow.

Note that only composition spaces $\KHS^\pp{\nu}$ with $[\nu]=[1^N]$  are isomorphic the permutation module $M^{[1^N]}$ and contain a totally antisymmetric irrep of $\mathrm{P}_N$. Therefore, only for compositions  of $N$ distinct states  does $\KHS^\pp{\nu}$ have an energy level that can be populated by one-component fermions. Because of this spatial antisymmetry, compositions with $[\nu]=[1^N]$ play an important role in the case of contact interactions, as discussed below.

\subsubsection{Bosonic and Fermionic Spectral Isomorphism}

Select the unique sequence $\tilde{\bf n}$ in the composition $\pp{\nu}$ with state labels arranged in increasing order. For example, this is the sequence $\tilde{\bf n}= \langle 0,1,1,2,4 \rangle$ from $\pp{\nu}=\pp{01^224}$. By adding $0$ to the first element of the sequence, $1$ to the next element and so on, a new composition $(\nu')$ is produced which has the shape $[\nu']=[1^N]$, e.g.\ the sequence $\tilde{\bf n}= \langle 0,1,1,2,4 \rangle$ with shape $[21^3]$ becomes the sequence $\tilde{\bf n}' = \langle 0,2,3,5,8 \rangle$ which has shape $[1^5]$. This means that for every totally symmetric state in $\KHS^\pp{\nu}$ (because there is always one copy of $\mathcal{M}^{[N]}$ in every composition subspace), there is a partner antisymmetric state in  $\KHS^\pp{\nu'}$, and the relationship is one-to-one. This mapping is a generalization of the result of Crescimanno~\cite{Crescimanno} that the bosonic and fermionic non-interacting spectrum have the same structure for the harmonic oscillator, just shifted by a constant value. It is also related to the famous Girardeau fermionization mapping of identical one-component bosons~\cite{Girardeau1960}, as discussed in Section 5.

\subsubsection{The Group $\mathrm{K}_N^0$ for Symmetric Traps}

For the general asymmetric trap, the kinematic symmetry is the minimal symmetry $\mathrm{K}_N^0 = \mathrm{P}_N \ltimes \mathrm{T}_t^{\times N}$ and irreps are labeled by compositions $\pp{\nu}$. These irreps sort into equivalence classes irreps based on their shapes $[\nu]$. Symmetric traps have the larger one-particle kinematic symmetry group $\mathrm{K}_1 = \mathrm{O}(1)\times \mathrm{T}_t$. Irreps are still characterized by compositions, but parity adds an additional quantum number. Even though two compositions have the same shape, they may not transform the same way under the single-particle parities $\hat{\Pi}_i$. As a result, not all $\mathcal{K}^\pp{\nu}$ subspaces with the same composition shape $[\nu]$ transform fall into the same equivalence class. For four particles, the possible equivalence class are (organized by shape)
\begin{eqnarray*}
{[4]}: && \pp{+^4}, \pp{-^4} \\
{[31]}: && \pp{+^3_1 +_2}, \pp{+^3 -}, \pp{-^3 +}, \pp{-^3_1 -_2}\\
{[2^2]}: && \pp{+^2_1 +^2_2}, \pp{+^2 -^2}, \pp{-^2_1 -^2_2}\\
{[21^2]}: && \pp{+^2_1 +_2 +_3}, \pp{+^2_1 +_2 -}, \pp{+^2 -_1 -_2}, \pp{-^2 +_1 +_2}, \pp{-^2_1 -_2 + }, \pp{-^2_1 -_2 -_3}\\
{[1^4]}: && \pp{+_1 +_2 +_3 +_4}, \pp{+_1 +_2 +_3 -}, \pp{+_1 +_2 -_1 -_2}, \pp{+_1 -_1 -_2 -_3}, \pp{-_1 -_2 -_3 -_4}.
\end{eqnarray*}
The notation $\pp{-^2 +_1 +_2}$ identifies an equivalence class of compositions with three states, one with negative parity and two distinct positive parity states, for example.

The group  $\mathrm{K}_1^{\times N}$ is an abelian normal subgroup of $\mathrm{K}_N^0$, and so the induced representation construction described at the end of Section~\ref{sec:induced} can also be used to identify these equivalence classes for $\mathrm{S}_N \ltimes (\mathrm{O}(1) \times \mathrm{T}_t)^{\times N}$. The reduction of composition subspaces $\KHS^\pp{\nu}$ into $\mathrm{S}_N$ irreps is unchanged. Note that the total number of distinct irreps of the minimally-constructed groups $\mathrm{C}_N^0$ and $\mathrm{K}_N^0$ are the same although the distribution of irrep dimensions is not. 

\subsubsection{Emergent and Accidental Symmetries}

When $\mathrm{K}_N^0$ is equal to the minimal kinematic symmetry group $\mathrm{P}_N \ltimes \mathrm{K}_1^{\times N}$, each composition subspace $\KHS^\pp{\nu}$ is uniquely related to an energy level $E_\pp{\nu}\in\sigma_N^0$. The reduction of the spatial Hilbert space into irreps of $\mathrm{K}_N^0$ is the same as the decomposition into composition subspaces (\ref{eq:compdecomp}). However, if there are coincident energy levels, i.e.\ two or more compositions lead to the same total energy, then there are additional kinematic symmetries.

The first case is  accidental degeneracies, like the Pythagorean degeneracies that occur in the infinite square well and its higher dimensional generalizations~\cite{Fernandez, Leyvraz}. These accidental degeneracies can be formulated as an ad hoc kinematic symmetry by defining operators that act as the identity in most energy subspaces but act unitarily in the accidentally-degenerate composition subspaces. In such a formulation, each accidental degeneracy requires the addition of new operators that commute with the Hamiltonian. Thinking of this as a kinematic symmetry is therefore not productive because these symmetry operators  must be inferred from the degeneracies and not the other way around. Accidental degeneracies of this sort will not be considered further here.

The other reason for coincident energy levels is that there is an emergent few-body symmetry, i.e.\ a symmetry that cannot be generated by the one-particle symmetries and particle permutations. The harmonic well is the most famous example. Its energy levels have a degeneracy larger than can be explained by $\mathrm{P}_N \ltimes \mathrm{K}_1^{\times N}$. For the energy level $\hbar\omega(X + N/2)$ with total excitation $X= \sum_i n_i$, the degeneracy is~\cite{Baker, Louck}
\begin{equation}
d(\mathrm{U}(N),X) = \frac{(X+N-1)!}{X!(N-1)!}.
\end{equation}
This degeneracy can be derived from combinatorics, or it can be explained by the fact that $\mathrm{K}_0^N \sim \mathrm{U}(N)$, the group of unitary transformations in $N$ dimensions. The $\mathrm{U}(N)$ symmetry can be thought of either as the group of symplectic, orthogonal transformations in phase space or as the group of unitary transformations of the $N$ annihilation operators: $\mathrm{Sp}(2N)\cap\mathrm{O}(2N) \sim \mathrm{U}(N)$. The irreps of $\mathrm{U}(N)$ can be labeled by total excitation $X$ and the irrep space is the direct sum of all composition spaces with the same total excitation.

\section{Interacting Particles}

The introduction of Galilean-invariant two-body interactions $\hat{V}_{ij}$ among the identical particles breaks the symmetry encapsulated by the subgroups $\mathrm{C}_1^{\times N}$ and $\mathrm{K}_1^{\times N}$.  The levels in the non-interacting energy spectrum $\sigma^0_N$ therefore split and degeneracies are reduced when two-body interactions are turned on.  However, the permutation symmetry subgroup is preserved, as well as any transformation in $\mathrm{C}_1^{\times N}$ or $\mathrm{K}_1^{\times N}$ that also commutes with the interaction operator
\begin{equation}
\hat{V}^N = \sum_{i<j}^N \hat{V}_{ij}.
\end{equation}
Denote the symmetry groups of $\hat{H}^N = \hat{H}^N_0 + \hat{V}^N$ by $\mathrm{C}_N$ and $\mathrm{K}_N$. They are subgroups of $\mathrm{C}_N^0$ and $\mathrm{K}_N^0$, respectively. See Table \ref{tab:lowNgrV} for information about $\mathrm{C}_N$ and $\mathrm{K}_N$ and their irreps for low particle numbers. Before classifying them for the three types of traps, we take a brief detour into the symmetries of the two-body matrix elements.

\begin{table}
\caption{This table provides information about the kinematic symmetry group $\mathrm{K}_N$ and configuration space symmetry group $\mathrm{C}_N$ for $N=2$, $3$ and $4$ identical non-interacting particles in asymmetric, symmetry and harmonic traps. For each group, the dimensions of the unitary irreducible representations (irreps) is provided, and the number of inequivalent irreps with that dimension is indicated by the superscript. The Coxeter notation and order is given for the finite configuration space groups, and the point group notation is also given for $N=2$ and $3$.}
\centering
\label{tab:lowNgrV}
\begin{tabular}{|c|c|c|c|}
\hline
 & Asym. Trap &  Sym. Trap &  Harm. Trap\\
\hline\hline\hline
$\mathrm{C}_2$ & $\mathrm{S}_2 $ & $\mathrm{S}_2\!\! \times\!\! \mathrm{O}(1) $ & $\mathrm{S}_2\!\! \times\!\! \mathrm{O}(1)$\\
\hline
Irreps & $1^2$ & $1^4$ & $1^4$ \\
\hline
Point & $\mathrm{D}_1$ & $\mathrm{D}_2$ & $\mathrm{D}_2$  \\
\hline
Coxeter & $\mathrm{A}_1\sim [\,]$ & $\mathrm{A}_1{}^2 \sim [2]$ & $\mathrm{A}_1{}^2 \sim [2]$ \\
\hline
 Order & $2$ & $4$ & $4$ \\
\hline\hline
$\mathrm{K}_2$  & $\mathrm{S}_2 \!\!\times\!\! \mathrm{T}_t$ & $\mathrm{S}_2 \!\!\times \!\!\mathrm{O}(1)\!\! \times \mathrm{T}_t$ & $\mathrm{S}_2\!\! \times\!\! \mathrm{U}(1)\!\!\times \!\!\mathrm{T}_t$ \\
\hline
Irreps & $1^2$ & $1^4$ & $1^4$\\
\hline\hline\hline

$\mathrm{C}_3$ & $\mathrm{S}_3 $ & $\mathrm{S}_3\!\! \times\!\! \mathrm{O}(1) $ & $\mathrm{S}_3 \!\!\times\!\! \mathrm{O}(1)^{\times 2}$\\
\hline
Irreps & $1^2,2$ & $1^4,2^2$ & $1^8,2^4$ \\
\hline
Point & $\mathrm{C}_{3v}$ & $\mathrm{D}_{3d}$ & $\mathrm{D}_{6h}$  \\
\hline
Coxeter & $\mathrm{A}_2\sim [3]$ & $[[3]]$ & $[[3],2]$ \\
\hline
 Order & $6$ & $12$ & $24$ \\
\hline
\hline
$\mathrm{K}_3$  & $\mathrm{S}_3\!\! \times\!\! \mathrm{T}_t$ & $\mathrm{S}_3 \!\!\times\!\! \mathrm{O}(1)\!\! \times \mathrm{T}_t$ & $\mathrm{S}_3 \!\!\times\!\! \mathrm{O}(1)\!\! \times\!\! \mathrm{U}(1)\!\!\times\!\! \mathrm{T}_t$ \\
\hline
Irreps & $1^2,2$ & $1^4,2^2$ & $1^8,2^4$\\
\hline
\hline\hline


$\mathrm{C}_4$ & $\mathrm{S}_4 $ & $\mathrm{S}_4 \!\!\times\!\! \mathrm{O}(1) $ & $\mathrm{S}_4 \!\!\times\!\! \mathrm{O}(1)^{\times 2}$\\
\hline
Irreps & $1^2,2,3^2$ & $1^4,2^2,3^4$ & $1^8,2^4,3^8$ \\
\hline
Coxeter & $\mathrm{A}_3\sim [3,3]$ & $[[3,3]]$ & $[[3,3],2]$ \\
\hline
 Order & $24$ & $48$ & $96$ \\
 \hline\hline
 $\mathrm{K}_4$  & $\mathrm{S}_4 \!\!\times\!\! \mathrm{T}_t$ & $\mathrm{S}_4 \!\!\times\!\! \mathrm{O}(1)\!\! \times\!\! \mathrm{T}_t$ & $\mathrm{S}_4\!\! \times \!\!\mathrm{O}(1) \!\!\times\!\! \mathrm{U}(1)\!\!\times\!\! \mathrm{T}_t$ \\
\hline
Irreps & $1^2,2,3^2$ & $1^4,2^2,3^4$ & $1^8,2^4,3^8$\\
\hline
\end{tabular}
\end{table}

\subsection{Symmetries of the Two-Body Matrix Elements}

The two-body matrix elements are the matrix elements of the two-body interaction in the non-interacting particle basis:
\begin{equation}\label{eq:v12}
\br{{\bf n}}\hat{V}_{12} \kt{{\bf n}'} = \br{n_1 n_2}\hat{V}_{12} \kt{n'_1 n'_2} \delta_{n_3 n'_3} \cdots \delta_{n_N n'_N} = v^{n_1 n_2}_{n'_1 n'_2} \delta_{n_3 n'_3} \cdots \delta_{n_N n'_N}
\end{equation}
Because of the hermiticity of $\hat{V}_{ij}$, and remembering that the stationary states of single trapped particles can always be chosen as real, the two-body matrix elements are also real and have the property
\begin{equation}
v^{\alpha\beta}_{\gamma\delta} = v_{\alpha\beta}^{\gamma\delta}.
\end{equation}
Galilean invariance constrains the position representation of the two-body interaction to have the form
\begin{equation}
\br{\bf q}\hat{V}_{ij} \kt{{\bf q}'}= V^2(|q_i-q_j|)\delta^N({\bf q} - {\bf q}').
\end{equation}
where $V^2$ is a scalar function of particle separation. The potential $V^2$ could be attractive or repulse, weak or strong, short range or long range; all that matters is that it is Galilean invariant.
One consequence of this invariance is $\hat{V}_{ij}=\hat{V}_{ji}$, so the two-body matrix elements also have the property
\begin{equation}
v^{\alpha\beta}_{\gamma\delta} = v^{\beta\alpha}_{\delta\gamma}.
\end{equation}
Putting these together, the following four two-body matrix elements are equivalent for any Galilean-invariant two-body interaction potential $\hat{V}_{ij}$:
\begin{equation}\label{eq:gime}
v^{\alpha\beta}_{\gamma\delta} = v_{\alpha\beta}^{\gamma\delta} = v^{\beta\alpha}_{\delta\gamma} = v_{\beta\alpha}^{\delta\gamma}.
\end{equation}
To better represent this symmetry, denote this two-body matrix element by $v_{\pp{\alpha\gamma}\pp{\beta\delta}}$, where the order or the symbols within the pair and the order of the pairs is arbitrary. This matrix element is proportional to the direct, first order transition amplitude for state transitions $\alpha\leftrightarrow\gamma$ and $\beta\leftrightarrow\delta$.

Additionally, for the contact interaction the two-body matrix elements also have the property
\begin{equation}\label{eq:cime}
v^{\alpha\beta}_{\gamma\delta} = v^{\alpha\beta}_{\delta\gamma} = v^{\beta\alpha}_{\gamma\delta}
\end{equation}
because
\begin{eqnarray}
\br{\alpha\beta}\hat{V}_{12} \kt{\gamma\delta} &=& g \int dq_1dq_2 \phi_\alpha^*(q_1) \phi_\beta^*(q_2) \delta(q_1 -q_2)  \phi_\gamma(q_1) \phi_\delta(q_2)  \nonumber\\
&=& g \int dq \phi_\alpha(q) \phi_\beta(q) \phi_\gamma(q) \phi_\delta(q).
\end{eqnarray}
Combining (\ref{eq:gime}) and (\ref{eq:cime}), for contact interactions the two-body matrix elements $v^{\alpha\beta}_{\gamma\zeta}$ with all of the 24 possible permutations of the state labels take the same value. Therefore, for the contact interaction the notation $v_\pp{\alpha\beta\gamma\delta}$ is convenient. This state permutation invariance of the contact interaction is not shared by the Hamiltonians $\hat{H}^N_0$ or $\hat{H}^N$ but it provides an alternate explanation for why the states in totally antisymmetric representations of $\mathrm{P}_N$ are unperturbed by the contact interaction. Since  $\hat{V}^N$ is symmetric under exchange of particles and the two-body matrix elements $v_\pp{\alpha\beta\gamma\delta}$ are symmetric under exchange of states, $\hat{V}^N$ annihilates states in the antisymmetric subspace $\KHS^{[1^N]}$.

\subsection{Configuration Space Symmetry Group $\mathrm{C}_N$}

Any Galilean-invariant operator $\hat{V}^N$ has the same configuration space symmetry as the coincidence manifold  $\mathcal{V}^N$ of all configurations in which at least two particles coincide in position~\cite{Brouzos2012a}
\begin{equation}
\mathcal{V}^N = \bigcup_{i<j}^N \mathcal{V}_{ij},
\end{equation}
where $\mathcal{V}_{ij}$ is the $(N\!-\!1)$-dimensional hyperplane with $q_i=q_j$. The coincidence manifold divides configuration space  $\mathcal{Q}^N$ into $N!$ identical sectors\footnote{This is an essential difference between one dimension and higher dimensions: in higher dimensions the particles can slip past each other without the configuration passing through the coincidence manifold $\mathcal{V}^N$.}. By Galilean invariance, the operator $\hat{V}^N$ and manifold $\mathcal{V}^N$ are invariant under permutation of particles and under total inversion $\hat{\Pi}$. Additionally, Galilean invariance implies that $\hat{V}^N$ commutes with the total momentum $\hat{P} = \sum \hat{P}_i$, and therefore $\mathcal{V}^N$ is invariant under translations in (or inversions of) the  center-of-mass coordinate $\hat{Q} \propto \sum \hat{Q}_i $. The configuration space symmetry of the coincidence manifold $\mathcal{V}^N$ is therefore isomorphic to
\begin{equation}\label{eq:cv}
\mathrm{C}_N^V \sim \mathrm{S}_N \times \mathrm{O}(1) \times (\mathrm{O}(1) \ltimes \mathrm{T}_R ),
\end{equation}
where $\mathrm{T}_R$ are translations along the center-of-mass axis and the second copy of $\mathrm{O}(1)$ is reflections $\Pi_R$ perpendicular to the center-of-mass axis. The other copy of $\mathrm{O}(1)$ is total inversion $\Pi$. Relative inversion, i.e.\ inversion of the relative coordinates but preserving the orientation of the center-of-mass, is defined $\Pi_r \equiv \Pi_R \Pi= \Pi \Pi_R$ and is also in the group $\mathrm{C}_N^V$.

\begin{figure}
\centering
\includegraphics[width=.7\linewidth]{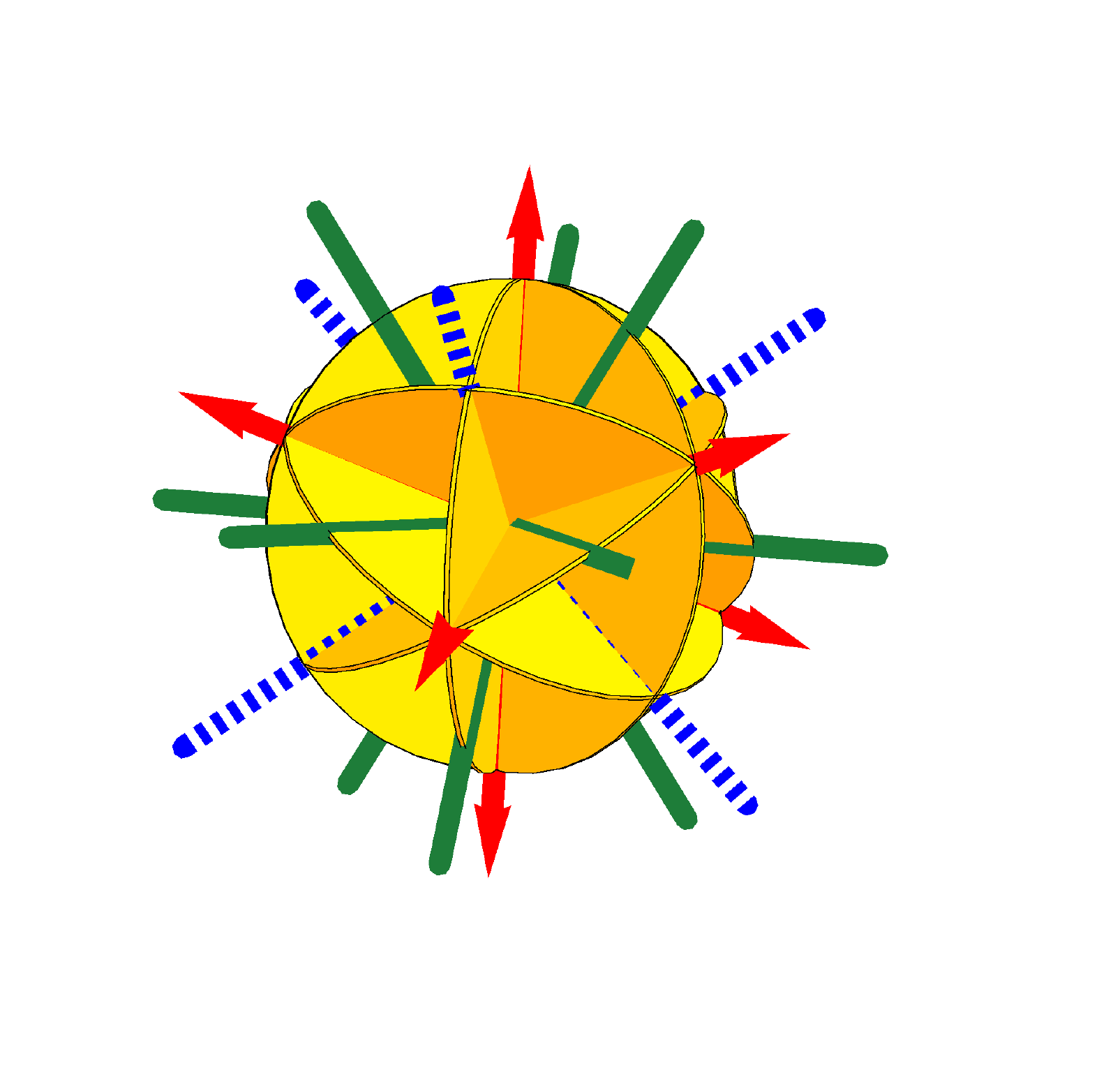}
\caption{This figure depicts the intersection of the four-dimensional, four-particle coincidence manifold $\mathcal{V}^4$ (the six disk planes) with a three-dimensional hyperplane of constant center-of-mass $q_1+q_2+q_3+q_4 = c$, i.e.\ the intersection of octahedral prism $\mathcal{V}^4$ and the relative configuration space. There are 24 sectors $\mathcal{Q}_s$ for each $s\in\mathrm{S}_4$ corresponding to a specific particle order $q_{s_1} < q_{s_2} < q_{s_3} < q_{s_4}$. Reflections across the disk planes are the geometrical realizations of transpositions like $(12)$. Three-cycles like $(123)$ correspond to rotations by $2\pi/3$ around the arrows (red online). Rotations by $\pi$ around the dashed line segments (blue online) correspond to two two-cycles like $(12)(34)$. Rotoreflections by $\pi/2$ around the solid line segments (green online) are four-cycles like $(1234)$.}
\label{fig:fourpart}
\end{figure}

Excluding the translation symmetry (which will be broken by any trap) the remaining symmetry group $\mathrm{S}_N \times \mathrm{O}(1)^{\times 2}$ is isomorphic to the point group of a $N$-dimensional prism with end faces that are dual $N$-simplices, e.g.\ a hexagonal prism for $N=3$ (see previous article) or an octahedral prism for $N=4$ (see Fig.~\ref{fig:fourpart}).

The configuration space symmetry group $\mathrm{C}_N$ of the total interacting Hamiltonian $\hat{H}^N$ must contain the intersection of the transformations in $\mathrm{C}^V_N$ and $\mathrm{C}^0_N$. Consider the three cases:
\begin{itemize}
\item Asymmetric well: $\mathrm{C}_N \sim \mathrm{S}_N$. The interacting Hamiltonian has the same configuration space symmetry as the non-interacting Hamiltonian, namely just $\mathrm{S}_N$. As described above, particle exchanges are realized in $\mathcal{Q}^N$ by orthogonal transformations that leave the center-of-mass axis and orientation invariant.
\item Symmetric well: $\mathrm{C}_N \sim \mathrm{S}_N \times \mathrm{O}(1)$. Total inversion $\hat{\Pi}$ is a symmetry transformation and the $2N!$ elements of $\mathrm{C}_N$ are realized by orthogonal transformations that leave the center-of-mass axis (but not necessarily the orientation) invariant. There is a doubling of irreps compared to the previous case, one set of $[\mu]\in P(N)$ with even parity $[\mu]^+$ and one set with odd parity $[\mu]^-$. Note that even for symmetric traps relative parity is not a good quantum number for $N>2$ interacting particles  unless the external single-particle potential $V^1(\hat{Q}_i)$ is quadratic in position.
\item Harmonic well: $\mathrm{C}_N \sim \mathrm{S}_N \times \mathrm{O}(1)^{\times 2}$. For quadratic trapping potentials like the harmonic well, the center-of-mass and relative coordinates are separable.  The total inversion $\Pi$ and center-of-mass inversion $\Pi_R$ (or equivalently $\Pi_R$ and relative inversion $\Pi_r$) are independently good symmetry transformations.  The group $\mathrm{S}_N \times \mathrm{O}(1)^{\times 2}$ has order $4N!$. The four irreps $[\mu]^\pm_\pm$ for every partition $[\mu]\in P(N)$ correspond to the four possible combinations of total and relative parity. Note that for the harmonic well, compositions $\pp{\nu}$ of non-interacting states have definite total parity but such compositions need not have definite relative parity (or center-of-mass quantum number, see below). There are multiple compositions with the same energy for the harmonic well, there are other observables that better interpolate between the $\mathrm{U}(N)$ symmetry of the non-interacting case and the $\mathrm{S}_N \times \mathrm{O}(1)^{\times 2}$ symmetry of the interacting case, namely the Jacobi hypercylindrical basis. For more details, see Ref.~\cite{Harshman2014}.
\end{itemize}

\subsection{Kinematic Symmetry Group $\mathrm{K}_N$}

Unless there are accidental or emergent symmetries, the kinematic symmetry group for interacting particles is just $\mathrm{P}_N \times \mathrm{K}_1 \sim \mathrm{C}_N \times \mathrm{T}_t$, where in this case the time translation group $\mathrm{T}_t$ generated by $\hat{H}^N$ is just a phase in each energy eigenspace. The clocks of the individual particles are synchronized by the interaction and are no longer independent symmetries. For asymmetric traps and symmetric (non-harmonic) traps, the irrep structure of $\mathrm{K}_N$ differs from $\mathrm{C}_N$ only by the addition of an extra quantum number, the total energy.

For the harmonic trap, there is the additional kinematic symmetry factor $\mathrm{U}(1)$ due to the separability of center-of-mass and relative coordinates. The center-of-mass degree of freedom in phase space is unperturbed by the interaction. The kinematic symmetry for the harmonic potential is therefore $\mathrm{K}_N \sim  \mathrm{S}_N\! \times \!\mathrm{O}(1) \!\times\! \mathrm{U}(1)\!\times\! \mathrm{T}_t$. The $\mathrm{U}(1)$ symmetry gives each irrep of this group an additional quantum number $n$. There are an infinite number of irreps of $\mathrm{K}_N$ labeled $[\mu]^\pm_n$, none of mix under interactions. For the case of contact interactions, this is exploited in Ref.\ \cite{Harshman2012, Garcia2014, Loft2014} to further decompose the kinematic Hilbert space into non-interacting subspaces that aid analysis and and efficient exact diagonalization and in Ref.\ \cite{Harshman2014} to construct adiabatic mappings.

\subsection{Approximation Methods}

Analytic solutions of the model Hamiltonian are only known for a few trap shapes and interactions:
\begin{itemize}
\item harmonic traps with harmonic interactions\footnote{This system has kinematic symmetry $\mathrm{K}_N \sim \mathrm{U}(1)\times\mathrm{U}(N-1) \times \mathrm{T}_t^{\times 2}$. This is an example of an emergent interaction symmetry.};
\item harmonic traps with contact interactions of any strength for two particles~\cite{avakian_1987, busch_two_1998, jonsell_interaction_2002};
\item hard wall traps with contact interactions of any strength for $N$ particles (Bethe ansatz solution)~\cite{Gaudin1971, Gaudin, Oelkers2006, Hao2009}.
\end{itemize}
For almost everything else, we need approximation methods. Symmetry makes it easier to make these approximations in an analytically and computationally efficient way, and provides spectroscopic classifications that aid in the interpretation of the results.

\subsubsection{Exact diagonalization in a truncated Hilbert space}

As an example, consider exact diagonalization in the non-interacting basis. In Section 3, the energy eigenspaces $\KHS^\pp{\nu}$ of the non-interacting Hamiltonian $\hat{H}_0^N$ were shown to carry irreps of $\mathrm{K}^0_N$ and to be isomorphic to permutation modules $M^{[\nu]}$. The reduction of the the spaces $\KHS^\pp{\nu}$  into $\mathrm{S}_N$ irreps $\mrep{[\mu]}$ only depends on the shape of the composition $[\nu]$. This critical observation is best exploited in the double tableau basis $\kt{W\,Y}$ for $\KHS$. Because of the $\mathrm{S}_N$ symmetry, the matrix elements of the interaction $\hat{V}^N$ are diagonal in $Y$:
\begin{equation}\label{eq:redmat}
\br{W\,Y} \hat{V}^N \kt{W' \, Y'} = \rybr{W} \hat{V}^N \rykt{W'} \delta_{YY'},
\end{equation}
where $\rybr{W} \hat{V}^N \rykt{W'}$ is a reduced matrix element. This reduced matrix element is a linear combination of two-body matrix elements like $v_{\pp{\alpha\beta}\pp{\gamma\delta}}$ for generic two-body interactions or $v_\pp{\alpha\beta\gamma\delta}$ for contact interactions. If the coefficients $\bk{\bf n}{WY}$ are known then one calculates
\begin{equation}
\rybr{W} \hat{V}^N \rykt{W'} = \sum_{{\bf n}\in \pp{\nu_W},{\bf n}' \in \pp{\nu_{W'}}}  \bk{WY}{\bf n} \br{\bf n} \hat{V}^N \kt{{\bf n}'} \bk{{\bf n}'}{W'Y}.
\end{equation}
As an example, the reduced matrix element $\rybr{\tiny \young(\alpha\delta,\beta,\gamma)} \hat{V}^4 \rykt{\tiny \young(\alpha\gamma,\beta,\delta)}$ is 
\[
\rybr{\tiny \young(\alpha\delta,\beta,\gamma)} \hat{V}^4 \rykt{\tiny \young(\alpha\gamma,\beta,\delta)} = \frac{\sqrt{2}}{3} \left( v_\pp{\alpha^2\delta^2} + v_\pp{\beta^2\delta^2} - 
   2 v_\pp{\gamma^2\delta^2} \right).
\]
The explicit expressions for $\rybr{W} \hat{V}^N \rykt{W'}$ in terms of two-body matrix elements are sometimes unnecessary for calculating physical results; see \cite{Yurovsky2014,Yurovsky2015} for examples of evaluating sum rules incorporating similar symmetrization techniques.

Note that $Y=Y'$ implies that all tableaux have the same shape $[W]=[W']=[Y]=[Y']$, and so there are only matrix elements between vectors in the same tower subspace $\KHS^{[W]}$ of $\mathrm{S}_N$ irreps. This leads to two reductions in the computational scaling of exact diagonalization. First, one only needs to calculate matrix elements between states with the same shape $[W]$, and this means a reduction of necessary matrix elements by roughly a factor of the number of irreps for $N$. Second, for multi-dimensional irreps $[\mu]$, only one matrix element between basis vectors with the same Young tableaux $Y$ is necessary (all the rest are the same), providing another reduction for each sector  $\KHS^{[W]}$ of $d[W]$. 

To see how this works in a truncated Hilbert space, consider four particles in an asymmetric trap and compositions $\pp{\nu}$ such that
\[
0 \cdot\nu_0+1\cdot\nu_1+2\cdot\nu_2+3\cdot\nu_3+4\cdot\nu_4 \leq 4.
\]
There are twelve compositions in this truncation:
\[
\pp{0^4}, \pp{0^31}, \pp{0^32}, \pp{0^21^2}, \pp{0^33}, \pp{0^212}, \pp{01^3}, \pp{0^34}, \pp{0^213}, \pp{0^22^2}, \pp{01^22}, \pp{1^4}
\]
and there are 70 basis states in these twelve compositions\footnote{Note that except for the harmonic trap, truncating the Hilbert space only to these compositions may or may not be consistent from a total energy perspective, but it is still an illustrative example.}. However, for spinless bosons, only the twelve symmetric states with $[W]=[4]$ (one for each composition) are necessary for exact diagonalization in this truncated space. Spin-1/2 fermions could populate spatial states in subspaces $\KHS^W$ with $[W]=[2^2]$, $[21^2]$, $[1^4]$, but  there are no states with $[W]=[1^4]$ in this truncated Hilbert space. The ground state for these fermions must have spin-0 and therefore only five states  chosen from the $[2^2]$ subspace (different $W$ but all chosen with the same $Y$, e.g.\ $\tiny \young(12,34)$) are needed for exact diagonalization.

The larger $N$, the more irreps there are, and for a given truncation dimension, the greater the basis reduction provided by symmetrization. However, the dimension of the truncation necessary to achieve a target accuracy grows more rapidly with $N$ than the reduction due to symmetrization. So although symmetry methods reduce the computational challenge, the scaling of exact diagonalization remains a practical obstacle.

\subsubsection{Weak perturbations}

The same symmetry properties that facilitate exact diagonalization also assist degenerate perturbation theory for weak attractive or repulsive interactions. For example, unless there are emergent interaction symmetries, in first order perturbation theory each non-interacting subspace $\KHS^\pp{\nu}\sim M^{[\nu]}$ splits into some number of $\mathrm{S}_N$ irreps $[\mu]$ determined by the Kostka number $K_{[\nu][\mu]}$ (c.f.\ eq.~(\ref{eq:kostka})). For composition spaces $\KHS^\pp{\nu}$ that are simply reducible into $\mathrm{S}_N$ irreps, to first order in the weak perturbation the level with energy $E^\pp{\nu}$ splits into levels carrying the irrep $[\mu]$ for each $[\mu] \leq [\nu]$. The energy shifts depend on the trap shape and interaction, but the structure of level splitting and the first order energy eigenstates do not. For four particles, composition spaces like $\KHS^\pp{\alpha^4}$, $\KHS^\pp{\alpha^3\beta}$, and $\KHS^\pp{\alpha^2\beta^2}$ have this property.

However, for composition spaces that are not simply reducible, the interaction breaks state permutation symmetry and the Weyl tableaux $W$ no longer provide good quantum numbers for irreps that appear multiple times in the decomposition of  $\KHS^\pp{\nu}$. To find the first-order energy eigenvectors using degenerate perturbation theory means diagonalizing the matrix formed by the reduced matrix elements  $\rybr{W} \hat{V}^N \rykt{W'}$ defined in (\ref{eq:redmat}) for all $W$, $W'$ in a composition space $\KHS^\pp{\nu}$ with the same shape.

In the previous article, this done for $N=3$ for the mixed symmetry irrep $[21]$ in the non-simply reducible compositions with shape $[\nu]=[1^3]$. For  $N=4$, the compositions with shapes $[21^2]$ and $[1^4]$ are not simply reducible. 
For example, two copies of $\mrep{[31]}$ appear in $\KHS^\pp{\alpha^2\beta\gamma}$.
That means that in order to find the first order energy eigenvalues and eigenvectors of the two levels which carry the $\mathrm{S}_4$ irrep $[31]$ into which  $\KHS^\pp{\alpha^2\beta\gamma}$ splits, a two-by-two matrix must be diagonalized:
\[
\left( \begin{array}{cc}  \rybr{\tiny \young(\alpha\alpha\beta,\gamma)} \hat{V}^4 \rykt{\tiny \young(\alpha\alpha\beta,\gamma)} & \rybr{\tiny \young(\alpha\alpha\beta,\gamma)} \hat{V}^4 \rykt{\tiny \young(\alpha\alpha\gamma,\beta)} \\
\rybr{\tiny \young(\alpha\alpha\gamma,\beta)} \hat{V}^4 \rykt{\tiny \young(\alpha\alpha\beta,\gamma)}  & \rybr{\tiny \young(\alpha\alpha\gamma,\beta)} \hat{V}^4 \rykt{\tiny \young(\alpha\alpha\gamma,\beta)} \end{array} \right).
\]
Using expressions for these matrix elements for the contact interaction, these two three-dimensional levels have the energies
\begin{eqnarray}\label{eq:aabg}
&&\frac{1}{3}\left(2 v_\pp{\alpha^4} + 5 v_\pp{\alpha^2\beta^2} + 5 v_\pp{\alpha^2\gamma^2} + 
   2v_\pp{\beta^2\gamma^2} \right. \\
 && \left. {} \pm \sqrt{
   9 (v_\pp{\alpha^2\beta^2})^2 - 14 v_\pp{\alpha^2\beta^2} v_\pp{\alpha^2\gamma^2} + 
    9 (v_\pp{\alpha^2\gamma^2})^2 - 4 v_\pp{\alpha^2\beta^2} v_\pp{\beta^2\gamma^2} - 
    4 v_\pp{\alpha^2\gamma^2} v_\pp{\beta^2\gamma^2} + 4 (v_\pp{\beta^2\gamma^2})^2}\right)\nonumber
\end{eqnarray}
 This algebraic expression for the energy shift (and similar expressions for the first order energy eigenstates) manifests the weakest form of algebraic universality since the two-body matrix elements are explicitly required. For non-contact interactions, the matrix elements are more complicated and the algebraic expression  is longer than (\ref{eq:aabg}), but has the same form. To solve level splitting in levels like $\KHS^\pp{\alpha\beta\gamma^\delta}$, a three-by-three matrix must be diagonalized for irreps $[31]$ and $[21^2]$. That requires solving a cubic equation, so the algebraic expression are substantially longer, but still universal in this weakest sense.

For $N \geq 5$, the Kostka number $K_{[\nu][\mu]}$ for a given composition can be greater than four, and therefore expressions for the first order for some irreps require solving a characteristic equations for which we have no \emph{a priori} guarantee that a universal algebraic solution exists. While certainly there are algebraic solutions for some polynomials of order five and above (contact interactions in an infinite square well provides an example), I hypothesize that such algebraically-solvable high-order polynomials are idealized limiting cases and not typical for physically-realistic interacting systems.

If there are additional conserved quantities, like parity for symmetric traps or center-of-mass excitation for harmonic traps, then the spatial Hilbert space $\KHS$ can be further reduced into sectors that do not mix under interactions. As before, this deeper reduction allows more numerically efficient numerical solutions. For the case of parity, there is a doubling of the number of towers $\KHS^{[\mu]^\pm}$, one for each parity. However, there is no change to the results of first order perturbation theory because each composition space $\KHS^\pp{\nu}$ has a parity determined by its composition. For harmonic traps, the separability of center-of-mass and relative degrees of freedom makes the double tableau basis less useful, because state permutation symmetry does not commute with the unitary group $\mathrm{U}(1)$. However, this additional complexity is more than  compensated by the emergent kinematic symmetry $\mathrm{K}_N \sim  \mathrm{S}_N\! \times \!\mathrm{O}(1) \!\times\! \mathrm{U}(1)\!\times\! \mathrm{T}_t$. See \cite{Harshman2014} for some examples of how this scaling works for harmonic traps. Further techniques for maximally exploiting this additional symmetry for harmonic traps are the subject of an article currently under preparation.

\section{Unitary Limit of Contact Interactions}

In the unitary limit $g\rightarrow \infty$  of the contact interaction in one-dimension, the particles cannot get past each other. Classically, the particles rattle back and forth in the trap, bouncing with perfectly elastic collisions off each other or rebounding from the edge of the trap potential. Quantum mechanically, the scattering from the delta-potential is diffractionless~\cite{sutherland} and this system is called a Tonks-Girardeau gas.

The order of the particles is stable under these dynamics. The coincidence manifold $\mathcal{V}^N$ divides $\mathcal{Q}$ into $N!$ ordering sectors. Each order can be labels by a permutation $s\in\mathrm{S}_N$ and corresponds to a sector of configuration space $\mathcal{Q}_s\subset \mathcal{Q}^N$ defined by the condition $q_{s_1} < q_{s_2} < \cdots < q_{s_N}$. The sector $\mathcal{Q}_s$ is bounded by the $(N\!-\!1)$ hyperplanes $\mathcal{V}_{s_1s_2}$, $\mathcal{V}_{s_2s_3}$, \ldots, and $\mathcal{V}_{s_{N-1}s_N}$. See Fig.~ for a depiction of the four-particle coincidence manifold and sectors.

In the unitary limit of the contact interaction, the wave functions for finite energy states must vanish at the edges of the sectors $\mathcal{Q}_s$, but inside the sector they satisfy the non-interacting Hamiltonian $\hat{H}_0^N$. Denote by $L^2(\mathcal{Q}_s)$ the Hilbert space of wave functions on $\mathcal{Q}_s$ satisfying the nodal boundary conditions on $\mathcal{V}^N$. All the  ordering subspaces $\KHS_s\sim L^2(\mathcal{Q}_s)$ are equivalent. Define the spatial Hilbert subspace $\KHS_\infty$ by
\begin{equation}\label{eq:ordering}
\KHS_\infty=\bigoplus_{s\in\mathrm{S}_N} \KHS_s.
\end{equation}  
What about states in $\KHS$ but not in $\KHS_\infty$? They have infinite energy because they have some probability density on $\mathcal{V}^N$ or they have infinitely sharp discontinuities\footnote{Technically, one can construct normalized superpositions of finite-energy states in $\KHS_\infty$ that have expectation values for the total energy that diverge, but ignoring these kinds of pathological states I will refer to the space $\KHS_\infty$ as the spatial Hilbert space of finite energy states.}. Note that states in a super-Tonks-Girardeau gas (corresponding to the unbound states in the limit $g \rightarrow -\infty$) are also in $\KHS_\infty$, but the infinite tower of infinite negative-energy bounds states are not.

The energy spectrum within each these independent sectors $\KHS_s$ can be deduced from the one-particle spectrum $\sigma_1$ quite simply: whenever there is a fermionic state in the $N$-particle non-interacting spectrum $\sigma^0_N$, there is a stationary state in $\KHS_s$. This is because the boundary condition imposed by contact interactions is `automatically' solved by the non-interacting fermionic states due to antisymmetry. At the unitary limit, wave functions must vanish on the coincidence manifold $\mathcal{V}^N$. From a given one-particle spectrum $\sigma_1$, the spectrum $\sigma_N^\infty$ and many features of the states at the unitary limit can be determined universally, including the incorporation of identical particles with spin~\cite{Guan2009, Ma2009, Fang2011, Cui2014, Harshman2014, Girardeau2007, Deuretzbacher2008, Yang2009, Girardeau2010a, Girardeau2010b, Girardeau2011}. All ordering sectors $\KHS_s$ are identical and so the $N$-particle spectrum $\sigma_N^\infty$ in the unitary limit is composed of $N!$-degenerate energy levels.

\subsection{Configuration Space Symmetries}

The configuration space symmetry of the Hamiltonian $\hat{H}^N_\infty$ at the unitary limit must contain $\mathrm{C}_N$ as a subgroup because it is just a special case of a Galilean-invariant interaction. Therefore particle permutations $p\in\mathrm{P}_N$ are valid symmetries. The configuration space representation $\underline{O}(p)$ maps sectors $\mathcal{Q}_s$ onto each other like
\begin{equation}\label{sectrep}
\underline{O}(p)\mathcal{Q}_s = \mathcal{Q}_{sp^{-1}}.
\end{equation}
and the unitary representations map subspaces $\KHS_s$ onto each other like
\begin{equation}
\hat{U}(p)\KHS_s = \KHS_{sp^{-1}}.
\end{equation}
These transformations are  linear and continuous in $\mathcal{Q}^N$. However, there are other configuration space transformations in $\mathcal{Q}^N$ that are non-linear and non-continuous, but whose unitary representations on $\KHS_\infty$ commute with $\hat{H}^N_\infty$. Namely, any other map that permutes the sector labels $s$ like
\begin{equation}
\underline{M} s = s'
\end{equation}
is a configuration space symmetry. In this notation, particle permutations are represented by $N \times N$ matrices that act on the vector space of sector labels like
\[
\underline{M}(p) s = s p^{-1}
\]
and this is another example of the defining representation of $\mathrm{P}_N \sim \mathrm{S}_N$.

The set of all maps like $\underline{M}$ form a group $\mathrm{Q}_N$ that is isomorphic to the symmetric group $\mathrm{S}_{N!}$. The particle permutations $\mathrm{P}_N \subset \mathrm{Q}_N$ are the only such maps that have continuous representations on $\mathcal{Q}_N$. All the other maps $\underline{M} \in \mathrm{Q}_3$ are discontinuous, but because of the nodal structure  $\mathcal{V}^N$, the continuity of wave functions in $\KHS_\infty$ is not disrupted. 

Ordering permutations $\mathfrak{O}_N$ are a subgroup of sector permutations $\mathrm{Q}_N$ that are complementary to $\mathrm{P}_N$. Whereas $p \in \mathrm{P}_N$ exchange particle labels (i.e.\ the numbers) in a sector $s=\{s_1 s_2 \cdots s_N \}$ wherever they occur, ordering permutations $\mathfrak{o} \in \mathfrak{O}_N$ exchange the order of labels in $s$, no matter what number is in that place in the order. Ordering permutations are indicated by letters, e.g.\ $(ABC)$ means $s_1$ becomes $s_2$, $s_2$ becomes $s_3$, and $s_3$ becomes $s_1$, or in other words they act like normal permutations on the sectors
\[
\underline{M}(\mathfrak{o}) s = \mathfrak{o} s .
\]
As an example with four particles, the particle permutation $(12)$ exchanges the numbers 1 and 2 in each sector $s$, e.g.\ $\underline{M}(12) \{3412\} = \{3421\}$ and $\underline{M}(12) \{2431\} = \{1432\}$. In contrast, the ordering permutation $(AB)$ exchanges the first and second number in the permutation $s$, no matter what the numbers are, e.g.\  $\underline{M}(AB) \{3412\} = \{4312\}$ and $\underline{M}(AB) \{2431\} = \{4231\}$.
Therefore, in addition to carrying a representation of the normal particle permutation symmetry, $\KHS_\infty$ also carries a copy of ordering permutation symmetry $\mathfrak{O}_N$ that (like symbol permutation symmetry in permutation modules $M^{[1^N]}$) is isomorphic to $\mathrm{S}_N$. One way to denote the configuration space $\mathrm{C}_N^\infty$ symmetry group of $\hat{H}^N_\infty$  is therefore
\begin{equation}
\mathrm{C}_N^\infty \supset \mathrm{P}_N \times \mathfrak{O}_N \times \mathrm{C}_1 \sim \mathrm{S}_N^{\times 2} \times \mathrm{C}_1.
\end{equation}
A similar construction holds for $\mathrm{K}_N^\infty$.

The rest of this section applies these  dual symmetries of particle permutation and ordering permutation. The inclusion of parity and additional kinematic and dynamic symmetries is also discussed. The combination of these symmetries provide an alternate, complementary method for treating the near-unitary limits as those described in \cite{Deuretzbacher2014,Volosniev2013,Levinsen2014,Gharashi2015,Yang2015}.

\subsection{Snippet Basis}

One basis for $\KHS_\infty$ is provided by the snippet basis~\cite{Fang2011, Deuretzbacher2008, Deuretzbacher2014}.
Denote by $\kt{\pp{\nu}[1^N]}$ the single, totally antisymmetric state in a composition space $\KHS^\pp{\nu}$ that has composition shape $[\nu]=[1^N]$. Then the snippet basis vectors are the $N!$ states denoted $\kt{\pp{\nu}[1^N]; s}$ with the property
\begin{equation}
\bk{\bf q}{\pp{\nu}[1^N];s} = \left\{ \begin{array}{ll} \pi_s \sqrt{N!} \bk{\bf q}{\pp{\nu}[1^N]} & {\bf q} \in \mathcal{Q}_s \\
0 & {\bf q} \ni \mathcal{Q}_s \end{array} \right.
\end{equation}
Each spatial Hilbert space sector $\KHS_s$ is spanned by the infinite tower of states $\kt{\pp{\nu}[1^N];s}$ for all $[\nu] = [1^N]$. The $N!$ snippet basis vectors $\kt{\pp{\nu}[1^N];s}$ span an $N!$-degenerate energy level $\KHS_\infty^\pp{\nu}$ with energy $E_\pp{\nu}$.

Sector permutations $\underline{M}\in\mathrm{Q}_3$ transform snippet vectors as expected
\[
\hat{U}(\underline{M})\kt{(\nu)[1^N];s} = \kt{(\nu)[1^N]; \underline{M} s}
\]
and specifically particle permutations $p\in\mathrm{P}_N$ and ordering permutations $\mathfrak{o}\in\mathfrak{O}_N$ 
have the representations
\begin{eqnarray}\label{sniprep}
\hat{U}(p)\kt{\pp{\nu}[1^N];s} &=& \kt{\pp{\nu}[1^N]; s p^{-1}}\ \mbox{and} \nonumber\\
\hat{U}(\mathfrak{o})\kt{\pp{\nu}[1^N];s} &=& \kt{\pp{\nu}[1^N]; \mathfrak{o}s }.
\end{eqnarray}
From this representation it is clear that particle permutations and ordering permutations commute.

The representations (\ref{sniprep}) of both copies of $\mathrm{S}_N$ are $N!$-dimensional and therefore must be reducible. In fact they are both the regular representations of $\mathrm{S}_N$ and the representation space $\mathrm{K}^\pp{\nu}_\infty$ is isomorphic to the permutation module $M^{[1^N]}$.
As with state permutation symmetry in compositions spaces $\KHS^\pp{\nu}$ with $[\nu]=[1^N]$, the canonical subgroup chains $\mathrm{S}_N \supset \mathrm{S}_{N-1} \supset \cdots\supset \mathrm{S}_2$ and $\mathfrak{O}_N \supset \mathfrak{O}_{N-1} \supset \cdots\supset \mathfrak{O}_2$ can be used to construct a complete set of commuting observables for the space $\mathrm{K}^\pp{\nu}_\infty$. The eigenvectors for these operators can be chosen as a double tableau basis $\kt{(\nu)[1^N];\mathfrak{Y}\,Y}$, where $\mathfrak{Y}$ is a Young tableau filled with symbols $A$, $B$, $C$, etc.\ that denote the irrep in which the ordering permutation subgroup chain is diagonalized and the Young tableau $Y$ filled with symbols $1$, $2$, $3$, etc. denotes the irrep in which the particle permutation subgroup chain is diagonalized (see also Ref.~\cite{Chen}). This notation was used for $N=3$ in the previous article. For $N=4$, the twenty-four basis vectors constructed from the fermionic state in $\KHS^\pp{\alpha\beta\gamma\delta}$ are
\begin{eqnarray*}
&&\ykt{\tiny \young(\alpha,\beta,\gamma,\delta) \normalsize ; \tiny \young(ABCD) \, \young(1234)} \equiv \ykt{\tiny \young(\alpha,\beta,\gamma,\delta) \normalsize ; \tiny \young(ABCD)},
\ykt{\tiny \young(\alpha,\beta,\gamma,\delta) \normalsize ; \tiny \young(ABC,D) \, \young(123,4)},
\ykt{\tiny \young(\alpha,\beta,\gamma,\delta) \normalsize ; \tiny \young(ABC,D) \, \young(124,3)},
\ykt{\tiny \young(\alpha,\beta,\gamma,\delta) \normalsize ; \tiny \young(ABC,D) \, \young(134,2)},\\
&&\ykt{\tiny \young(\alpha,\beta,\gamma,\delta) \normalsize ; \tiny \young(ABD,C) \, \young(123,4)},
\ykt{\tiny \young(\alpha,\beta,\gamma,\delta) \normalsize ; \tiny \young(ABD,C) \, \young(124,3)},
\ykt{\tiny \young(\alpha,\beta,\gamma,\delta) \normalsize ; \tiny \young(ABD,C) \, \young(134,2)},
\ykt{\tiny \young(\alpha,\beta,\gamma,\delta) \normalsize ; \tiny \young(ACD,B) \, \young(123,4)},
\ykt{\tiny \young(\alpha,\beta,\gamma,\delta) \normalsize ; \tiny \young(ACD,B) \, \young(124,3)},
\ykt{\tiny \young(\alpha,\beta,\gamma,\delta) \normalsize ; \tiny \young(ACD,B) \, \young(134,2)},\\
&&\ykt{\tiny \young(\alpha,\beta,\gamma,\delta) \normalsize ; \tiny \young(AB,CD) \, \young(12,34)},
\ykt{\tiny \young(\alpha,\beta,\gamma,\delta) \normalsize ; \tiny \young(AB,CD) \, \young(13,24)},
\ykt{\tiny \young(\alpha,\beta,\gamma,\delta) \normalsize ; \tiny \young(AC,BD) \, \young(12,34)},
\ykt{\tiny \young(\alpha,\beta,\gamma,\delta) \normalsize ; \tiny \young(AC,BD) \, \young(13,24)},\\
&&\ykt{\tiny \young(\alpha,\beta,\gamma,\delta) \normalsize ; \tiny \young(AB,C,D) \, \young(12,3,4)},
\ykt{\tiny \young(\alpha,\beta,\gamma,\delta) \normalsize ; \tiny \young(AB,C,D) \, \young(13,2,4)},
\ykt{\tiny \young(\alpha,\beta,\gamma,\delta) \normalsize ; \tiny \young(AB,C,D) \, \young(14,2,3)},
\ykt{\tiny \young(\alpha,\beta,\gamma,\delta) \normalsize ; \tiny \young(AC,B,D) \, \young(12,3,4)},
\ykt{\tiny \young(\alpha,\beta,\gamma,\delta) \normalsize ; \tiny \young(AC,B,D) \, \young(13,2,4)},
\ykt{\tiny \young(\alpha,\beta,\gamma,\delta) \normalsize ; \tiny \young(AC,B,D) \, \young(14,2,3)},\\
&&\ykt{\tiny \young(\alpha,\beta,\gamma,\delta) \normalsize ; \tiny \young(AD,B,C) \, \young(12,3,4)},
\ykt{\tiny \young(\alpha,\beta,\gamma,\delta) \normalsize ; \tiny \young(AD,B,C) \, \young(13,2,4)},
\ykt{\tiny \young(\alpha,\beta,\gamma,\delta) \normalsize ; \tiny \young(AD,B,C) \, \young(14,2,3)},
\ykt{\tiny \young(\alpha,\beta,\gamma,\delta) \normalsize ; \tiny \young(A,B,C,D) \, \young(1,2,3,4)}\equiv\ykt{\tiny \young(\alpha,\beta,\gamma,\delta)}.
\end{eqnarray*}
In this basis, the $\mathrm{S}_4$ irreps given by the shape $[\mathfrak{Y}]=[Y]$.  The coefficients between the snippet basis and the double tableau basis $\bk{\pp{\nu}[1^N];s}{\pp{\nu}[1^N];\mathfrak{Y}\,Y}$ can be found efficiently using computer-assisted algebra and the methods of Ref.~\cite{Chen}.

Note that every composition $\pp{\nu}$ with arbitrary shape $[\nu]$ is associated with another composition $\pp{\nu'}$ with shape $[1^N]$ by adding the sequence $\langle 0,1,2,\ldots, (\!N-\!1) \rangle$ to the representative ordered sequence $\tilde{\bf n}\in\pp{\nu}$. Therefore there is also an element $\kt{\pp{\nu'}[1^N];s}\in\KHS_p$ for every composition $\pp{\nu}$, i.e.\ the Bose-Fermi mapping~\cite{Girardeau1960}.

\subsection{Incorporating Additional Trap Symmetries}

One element $\mathfrak{o}_\pi \in \mathfrak{O}_N$ of the ordering permutation group is equivalent to reversing the order of the particles. For example, for three particles $\mathfrak{o}_\pi$ is $(AC)$ and for four particles $\mathfrak{o}_\pi=(AD)(BC) $. For symmetric wells, the parity operator is realized on the snippet basis using this operator:
\begin{eqnarray}
\hat{\Pi} \kt{\pp{\nu}[1^N];s} &=& \hat{\Pi} \kt{\pp{\nu}[1^N];\{s_1s_2 \cdots s_{N-1} s_N \}} \nonumber\\
&=& \pi_\pp{\nu} \hat{U}(\mathfrak{o}_\pi) \kt{\pp{\nu}[1^N]; s}\nonumber\\
&=& \pi_\pp{\nu} \kt{\pp{\nu}[1^N];\mathfrak{o}_\pi s}\nonumber\\
&=& \pi_\pp{\nu} \kt{\pp{\nu}[1^N]; \{s_N s_{N-1} \cdots s_2 s_1 \}},
\end{eqnarray}
where $\pi_\pp{\nu} = \pi_{\nu_1}\pi_{\nu_2}\cdots \pi_{\nu_r}$ is the total parity of the composition $\pp{\nu}$. This representation of the parity operator can be diagonalized to decompose $\KHS_\infty^\pp{\nu}$ into $\mathrm{S}_N \times \mathrm{O}(1)$ irreps $\mathcal{M}^{[\mu]^\pm}$. These will not necessarily be the same states constructed using the order permutation symmetry subgroup chain that leads to the double tableau basis because the operator $\mathfrak{o}_\pi$ is not part of the subgroup chain $\mathfrak{O}_N \supset \mathfrak{O}_{N-1} \supset \cdots \mathfrak{O}_2$. However, $\mathfrak{o}_\pi$ can still be diagonalized along with the normal particle permutation group subgroup chain, and an alternate ordering permutation subgroup chain can be found. For example, one can show that for four particles a complete set of commuting observables that respects parity is
\begin{eqnarray}\label{eq:csco4unit}
\hat{C}^\pp{1234}_2 &=& \hat{U}(12) + \hat{U}(13) + \hat{U}(23) + \hat{U}(14) + \hat{U}(24) + \hat{U}(34),\ 
\hat{C}^\pp{123}_2 =  \hat{U}(12) + \hat{U}(13) + \hat{U}(23), \nonumber\\
&& \ \hat{C}^\pp{12}_2 =  \hat{U}(12), \hat{U}(\mathfrak{o}_\pi)=\hat{U}((AD)(BC)), \ \mbox{and}\ \hat{C}^\pp{AD}_2 = \hat{U}(AD).
\end{eqnarray}
This notation for conjugacy class operators is a slight modification of Ref.~\cite{Chen}, and it is extensible to larger $N$.

Not all of the $\mathrm{P}_N \sim \mathrm{S}_N$ irreps have the same parity as the parity of the original fermionic composition. For example, the subspace  $\KHS_\infty^\pp{\alpha\beta\gamma\delta}$ with $\pi_\pp{\alpha\beta\gamma\delta}= 1$ reduces into the following $\mathrm{S}_4 \times \mathrm{O}(1)$ irreps
\begin{equation}\label{eq:4spaceunit}
\KHS_\infty^\pp{\alpha\beta\gamma\delta} \sim 
\mathcal{M}^{[4]^+} \oplus \mathcal{M}^{[31]^+} \oplus 2 \mathcal{M}^{[2^2]^+} \oplus \mathcal{M}^{[21^2]^+} \oplus \mathcal{M}^{[1^4]^+} \oplus 2 \mathcal{M}^{[31]^-} \oplus 2 \mathcal{M}^{[21^2]^-}.
\end{equation}
Looking at the dimensions $d[\mu]$, we see that half the states have positive parity and half have negative parity; this result holds for all $N$.

Besides the examples with $N=2$ and $N=3$ in the previous article, further examples of this diagonalization are discussed for the harmonic case in \cite{Harshman2014} and tables for $N=3$, $N=4$ and $N=5$ are provided. In the harmonic case, relative parity and the center-of-mass excitation are also good quantum numbers. They are not necessarily commensurate with composition subspaces, but they can be diagonalized simultaneously with the particle and ordering permutations symmetries. The extra symmetry means that the unitary limit in a harmonic trap is superintegrable, and therefore in the near unitary limit I conjecture the system is likely to resist thermalization longer than less symmetric traps.

\subsection{Near Unitary Limit}

In the near unitary limit, ordering permutation symmetry is broken because there is tunneling between adjacent sectors $\mathcal{Q}_s$. At unitarity, the kinematic symmetry of the Hamiltonian $\hat{H}_\infty^N$ is $\mathrm{K}^\infty_N \supset \mathrm{P}_N \times \mathfrak{O}_N \times \mathrm{K}_1$, but away from unitarity the symmetry is only $\mathrm{K}_N \sim \mathrm{P}_N \times \mathrm{K}_1$. Therefore, each energy level $\KHS^\pp{\nu}_\infty$ splits into the $\mathrm{P}_N$ irreps that compose it.

The tunneling can be represented by an operator that breaks the ordering permutation symmetry and `lowers the energy cost' of having a cusp on the coincidence manifold $\mathcal{V}^N$. Then near unitary the Hamiltonian is split perturbatively into $\hat{H}^N = \hat{H}_\infty^N + \hat{T}$. A suitable operator $\hat{T}$ has the form
\begin{equation}\label{tunn}
\hat{T} = -a_{AB} \hat{U}(AB) - a_{BC} \hat{U}(BC) - \cdots -  a_{MN} \hat{U}(MN) - (a_{AB} + a_{BC} + \cdots) \hat{U}(e),
\end{equation}
where the real, positive coefficients $a_{AB}$, $a_{BC}$, etc.\ are the tunneling amplitudes for adjacent sectors~\cite{Deuretzbacher2014}:
\begin{equation}
a_{ij} = \frac{N!}{g} \int_{\mathcal{Q}_e} d^N{\bf q} \, \delta(q_i - q_j) \left| \frac{\partial \bk{\bf q}{\pp{\nu}[1^N]}}{\partial q_i}\right|^2 .
\end{equation}
The last term in (\ref{tunn}) that is proportional to the identity renormalizes the energy shift so that the totally-antisymmetric spatial state feels no effect of the tunneling. If the trap is symmetric, then $a_{AB}=a_{MN}$, $a_{BC}=a_{LM}$, etc. For understanding level splitting and the formation of bands that depend on trap shape, the ratio of the tunneling amplitudes (and not the absolute scale) is important.

\begin{figure}
\centering
\includegraphics[width=\linewidth]{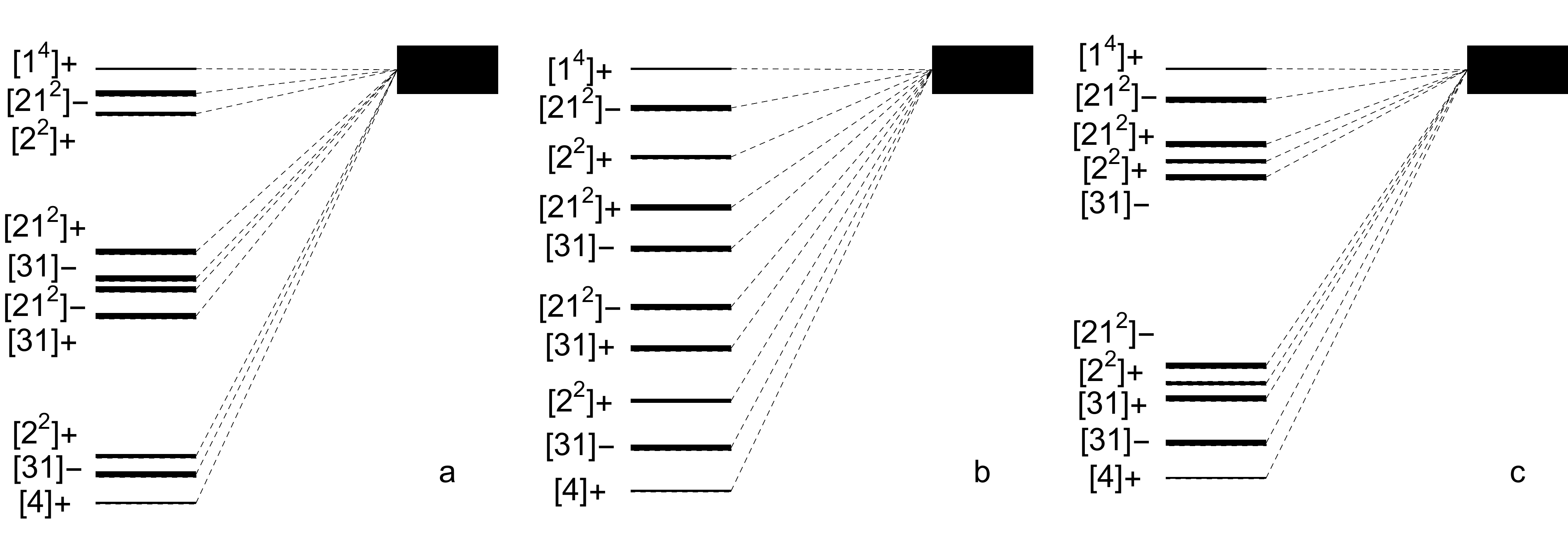}
\caption{Approximate level splitting diagram for four particles in three symmetric traps: (a) double well; (b) infinite square well; (c) V-shaped or cusped well, i.e. softer than harmonic. The thick band on the right in each figure is the 24-fold degenerate ground state energy level for four distinguishable particles in the unitary limit of the contact interaction. The three trap shapes are distinguished by the ratio of tunneling amplitudes $t/u$. The following ratios have been chosen to illustrate the trap dependence of these amplitudes: (a) $t/u=2.9$; (b) $t/u = 1$; (c) $t/u=0.3$. The idea is that (a) for double wells tunneling in the middle is suppressed so $t>u$; (b) for infinite square well the potential is uniform inside the trap and so (for low particle density) $t$ and $u$ are approximately the same; (c) for softer wells, there is more phase space in the middle of the well so $u>t$. For harmonic wells, $t/u \approx 0.762 $ (c.f.\ \cite{Deuretzbacher2014,Volosniev2013,Levinsen2014,Gharashi2015}). In subfigure (c) a more extreme ratio is depicted, corresponding to a V-shaped or cusped trap.}
\label{fig:level4}
\end{figure}

For example, see Fig.~\ref{fig:level4} (also reproduced in the introduction of the previous article). With four particles in a symmetric trap, the tunneling operator has the form
\begin{equation}
\hat{T} = -t \hat{U}(AB) - u \hat{U}(BC) - t \hat{U}(CD) - (2t + u) \hat{U}(e).
\end{equation}
Diagonalizing the basis provided by (\ref{eq:csco4unit}) shows that the energy splits into an energy level with $\pi_\pp{\nu} = 1$ for each $\mathrm{S}_4 \times \mathrm{O}(1)$ irrep in the sum (\ref{eq:4spaceunit}) as follows:
\begin{eqnarray}
{[4]}^+: && -4t - 2u \nonumber\\
{[31]}^+: && -2t-2u \nonumber\\
{[2^2]}^+: && -2t - u \pm \sqrt{4t^2 -2tu+u^2} \nonumber\\
{[21^2]}^+: && -2t \nonumber\\
{[1^4]}^+: && 0 \nonumber\\
{[31]}^-: && -3t - u \pm \sqrt{t^2 + u^2}\nonumber\\
{[21^2]}^-: && -t - u \pm \sqrt{t^2 + u^2}.
\end{eqnarray}
Unlike the three particle case, parity does not completely remove the trap dependence of the first order energy eigenstates; there are two copies of $[2^2]^+$, $[31]^-$, and $[21^2]^-$ that require an additional, trap dependent operator to diagonalize.

If instead an asymmetric trap was considered, there are three tunneling parameters for $\hat{U}(AB)$, $\hat{U}(BC)$ and $\hat{U}(CD)$. There are three copies of irreps $[31]$ and $[21^2]$, so computing level splitting requires solving a cubic equation. For $N = 5$ in asymmetric traps or $N=6$ in symmetric traps, the required diagonalizations require solving greater than quartic equations, and the last traces of algebraic universality are lost for the near-unitary regime. The characteristic polynomials have algebraic solutions for certain trap shapes (like the infinite square well), but as with the case of the weak level splitting, I hypothesize that those cases are not typical. 

When the level spitting problem is algebraically solvable in the weak interaction limit and the near unitary limit, then the adiabatic mapping problem between non-interacting states and their unitary limit is in principle solved. The order of the levels within a particle tower $\KHS^{[\mu]}$ should not change unless for some particular value of $g$ a new multiparticle symmetry emerges. That does not seem likely for the contact interaction, but I do not have a proof that it is impossible.

\section{Conclusion}

The essential claim of this pair of articles is that the configuration space and kinematic symmetries of $\hat{H}^N_0$, $\hat{H}^N$, and $\hat{H}^N_\infty$  provide a powerful and unifying tool for analysis and computation. One way to specify the focus of this article is to ask the series of questions:
\begin{itemize}
\item Given only the symmetries, what can we say about the $N$-particle spectrum $\sigma_N$? How much about the energies, degeneracies and states for the interacting system can be inferred without any specific knowledge of the one-particle spectrum $\sigma_1$? How does the spectrum change when the interaction strength is tuned adiabatically or rapidly quenched?
\item If we also know the specific spectrum $\sigma_1$ in addition to the symmetries, how much more can we say about the spectrum $\sigma_N$? And how much more if we also know the wave functions $\psi_n(q) = \bk{q}{n}$ of the one-particle energy eigenstates $\hat{H}^1 \kt{n} = \epsilon_n \kt{n}$?
\item Finally, what if we know the explicit form of the two-body interaction $\hat{V}_{ij}$ and/or can calculate the two-body matrix elements $\br{n_1 n_2} \hat{V}_{12} \kt{n_1' n_2'} = v_{\pp{n_1 n_2}\pp{n'_1 n'_2}}$ between noninteracting states? 
\end{itemize}
As a general conceptual framework, the less we have to know about the specifics of the trap or interaction, the more `universal' the results. What this article demonstrates is how this notion of universality breaks down for most interacting system as the number of particles is increased and the symmetries of the trap and interaction are reduced.

Despite its length, this article has left out many relevant topics, like the $\mathrm{SO}(2,1)$ dynamical symmetry of the contact potential in a harmonic trap, lattice symmetries, supersymmetric potentials in one dimensions, and the possibility of interaction symmetries that depend on the internal structure of the particles. Further, these symmetry classifications could be generalized to higher dimensions. Symmetry should be less constraining as the number of degrees of freedom grows and algebraic universality may break down sooner. Also, the effect of intrinsic three or higher  few-body interactions on the spectrum would be interesting to incorporate into symmetry analysis to seek possible manifestation in spectral shifts. 

However, without adding to the complication of the one-dimensional trap, two-body interaction model, there is still much work to be done. Efficient methods of state construction are required for perturbation theory and exact diagonalization and for calculating reduced density matrices, correlation functions, and entanglement spectra among particles and between spin and spatial observables. Another possible avenue for future work is to exploit the close connection between finite groups and number theory. Perhaps there are practical protocols for simulations of number theory problems that employ combinations of adiabatic tunings and diabatic quenches of the trap shape and and interaction strength to manipulate states.

\begin{acknowledgements}
The author would like to thank his students at American University who worked with him as performed this research in the last few years, especially J.~DeMell, J.~Hirtenstein, J.~Revels, M.~Roberts, A.~Taylor, J.~Verniero, and B.~Weinstein. He also humbly acknowledges the brain-extenders of Wikipedia and Mathematica.
\end{acknowledgements}

\end{document}